\documentclass[10pt,english,prb,amsmath,amssymb,prb,showkeys,superscriptaddress,twocolumn,showpacs,floatfix]{revtex4-1}
\usepackage{graphicx}
\usepackage{dcolumn}
\usepackage[colorlinks=true,linkcolor=blue ,citecolor=blue,urlcolor=blue]{hyperref}
\usepackage{multirow}
\usepackage[usenames,dvipsnames]{xcolor}
\usepackage{soul}
\usepackage{braket}
\usepackage{bm}
\usepackage{nicefrac}
\usepackage{siunitx}
\usepackage{color,transparent}
\usepackage[utf8]{inputenc}
\usepackage{upgreek}
\usepackage[export]{adjustbox}
\usepackage{pict2e}

\newcommand{\HH}{\mathcal{H}}

\newcommand{\VEC}[1]{\mathbf{#1}}

\DeclareSIUnit\ML{ML}
\DeclareSIUnit\MLs{MLs}
\DeclareSIUnit\meVA{meV\angstrom^2}

\begin{document}

\title{Complex magnetic structure and spin waves of the noncollinear antiferromagnet \texorpdfstring{\uppercase{M}\lowercase{n}\textsubscript{5}\uppercase{S}\lowercase{i}\textsubscript{3}}{Mn5Si3}} 

\author{N. Biniskos}
\email{n.biniskos@fz-juelich.de}
\affiliation{Forschungszentrum J\"ulich GmbH, J\"ulich Centre for Neutron Science at MLZ, Lichtenbergstr. 1, D-85748 Garching, Germany}
\author{F. J. dos Santos}
\email{flaviano.dossantos@epfl.ch}
\affiliation{Theory and Simulation of Materials (THEOS), and National Centre for Computational Design and Discovery of Novel Materials (MARVEL), \'Ecole Polytechnique F\'ed\'erale de Lausanne, 1015 Lausanne, Switzerland}
\author{K. Schmalzl}
\affiliation{Forschungszentrum J\"ulich GmbH, J\"ulich Centre for Neutron Science at ILL, 71 avenue des Martyrs, F-38000 Grenoble, France}
\author{S. Raymond}
\affiliation{Universit\'e Grenoble Alpes, CEA, IRIG, MEM, MDN, F-38000 Grenoble, France}
\author{M. dos Santos Dias}
\affiliation{Faculty of Physics, University of Duisburg-Essen and CENIDE, D-47053 Duisburg, Germany}
\affiliation{Peter Gr\"unberg Institut and Institute for Advanced Simulations, Forschungszentrum J\"ulich $\&$ JARA, D-52425 J\"ulich, Germany}
\author{J. Persson}
\affiliation{Forschungszentrum J\"ulich GmbH, J\"ulich Centre for Neutron Science (JCNS-2) and Peter Gr\"unberg Institut (PGI-4), JARA-FIT,  D-52425 J\"ulich, Germany}
\author{N. Marzari}
\affiliation{Theory and Simulation of Materials (THEOS), and National Centre for Computational Design and Discovery of Novel Materials (MARVEL), \'Ecole Polytechnique F\'ed\'erale de Lausanne, 1015 Lausanne, Switzerland}
\author{S. Bl\"ugel}
\affiliation{Peter Gr\"unberg Institut and Institute for Advanced Simulations, Forschungszentrum J\"ulich $\&$ JARA, D-52425 J\"ulich, Germany}
\author{S. Lounis}
\affiliation{Peter Gr\"unberg Institut and Institute for Advanced Simulations, Forschungszentrum J\"ulich $\&$ JARA, D-52425 J\"ulich, Germany}
\affiliation{Faculty of Physics, University of Duisburg-Essen and CENIDE, D-47053 Duisburg, Germany}
\author{T. Br\"uckel}
\affiliation{Forschungszentrum J\"ulich GmbH, J\"ulich Centre for Neutron Science (JCNS-2) and Peter Gr\"unberg Institut (PGI-4), JARA-FIT, D-52425 J\"ulich, Germany}

\date{\today}

\begin{abstract}
The investigations of the interconnection between micro-- and macroscopic properties of materials hosting noncollinear antiferromagnetic ground states are challenging.
These forefront studies are crucial for unraveling the underlying mechanisms at play, which may prove beneficial in designing cutting edge multifunctional materials for future applications.
In this context, $\textnormal{Mn}_5\textnormal{Si}_3$ has regained scientific interest since it displays an unusual and complex ground state, which is considered to be the origin of the anomalous transport and thermodynamic properties that it exhibits.
Here, we report the magnetic exchange couplings of the noncollinear antiferromagnetic phase of $\textnormal{Mn}_5\textnormal{Si}_3$ using inelastic neutron scattering measurements and density functional theory calculations.
We determine the ground-state spin configuration and compute its magnon dispersion relations which are in good agreement with the ones obtained experimentally.
Furthermore, we investigate the evolution of the spin texture under the application of an external magnetic field to demonstrate theoretically the multiple field-induced phase transitions that $\textnormal{Mn}_5\textnormal{Si}_3$ undergoes.
Finally, we model the stability of some of the material's magnetic moments under a magnetic field and we find that very susceptible magnetic moments in a frustrated arrangement can be tuned by the field.
\end{abstract}

\date{\today}

\maketitle

\section{Introduction}

The noncollinear spin arrangements in magnetic materials gives rise to important macroscopic phenomena that can be exploited for developing future information and communication technologies. 
Characteristic examples are the recent observation of large anomalous transport properties near room temperature, such as the anomalous Nernst~\cite{Ikhlas_2017} and Hall~\cite{Nakatsuji_2015,Kiyohara_2016} effects, in the metallic noncollinear antiferromagnetic systems $\textnormal{Mn}_3\textnormal{X}$ (with $\textnormal{X} = \textnormal{Sn}, \textnormal{Ge}$). 
In turn, these discoveries have initiated spectroscopic studies that have provided new insights of the intimate coupling between the various degrees of freedom, i.e., spin, lattice and electronic, which could explain the interesting anomalous phenomena in these materials~\cite{Park_2018,Sukhanov_2019,Chen_2020}. 
Therefore, it is evident that an experimental study of the spin dynamics and a comparison with theoretical models is crucial for understanding the origin of noncollinear spin arrangements and the peculiar properties that arise in ordered solid materials.

$\textnormal{Mn}_5\textnormal{Si}_3$ is another Mn-based metallic antiferromagnet (AFM) that has lately regained scientific interest owing to interesting thermodynamic (inverse magnetocaloric effect~\cite{songlin_magnetic_2002,biniskos_spin_2018}, inverted hysteresis loop, and thermomagnetic irreversibility~\cite{das_observation_2019}) and transport (large anomalous Hall effect~\cite{surgers_large_2014}) phenomena. It has two stable AFM phases, namely the AFM2 for $60 < T < 100$\,K and the AFM1 for $T < 60$\,K, that are confirmed by a plethora of macroscopic measurements~\cite{songlin_magnetic_2002,surgers_large_2014,Suergers_2016,surgers_switching_2017,das_observation_2019,luccas_magnetic_2019} in thin film, polycrystalline and single crystal samples. 
While neutron diffraction studies in powders~\cite{gottschilch_study_2012} and single crystals~\cite{brown_antiferromagnetism_1995} are in agreement regarding the collinear spin arrangement in the AFM2 phase, in the past years several contradicting spin structures~\cite{gottschilch_study_2012,Lander_1967,Menshikov_1990,brown_low-temperature_1992} have been proposed for the noncollinear AFM1 phase, where the interesting thermodynamic and transport properties are observed. So far, there has not been a dedicated study concerning the spin dynamics in the AFM1 phase, which could give additional insight in the magnetic spin structure. Therefore, in the present work, we perform inelastic neutron scattering (INS) measurements and density functional theory (DFT) calculations supplemented with various models to investigate spin waves in the noncollinear AFM1 phase of $\textnormal{Mn}_5\textnormal{Si}_3$ and to characterize the magnetic ground-state properties and electronic structure.

\begin{figure}[tb]
\setlength{\unitlength}{1cm} 
\newcommand{\boxsize}{0.3}
\begin{picture}(7,12.5)
    \put(0.0, 0.0){    
        \put(-0.4, 0.5){ \includegraphics[width=7.8cm,trim={0 0 0 0},clip=true]{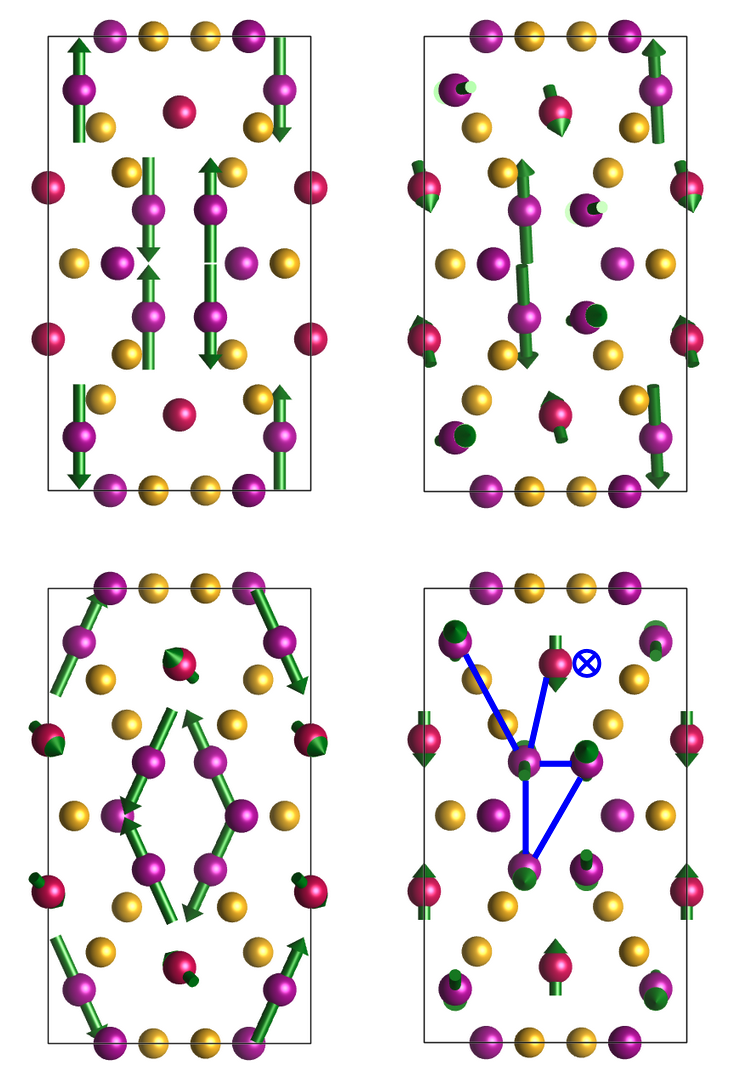} }
        
        \put(-0.5, 11.6){ \makebox(\boxsize,\boxsize){(a)} } 
        \put( 3.5, 11.6){ \makebox(\boxsize,\boxsize){(b)} } 
        \put(-0.5,  5.6){ \makebox(\boxsize,\boxsize){(c)} } 
        \put( 3.5,  5.6){ \makebox(\boxsize,\boxsize){(d)} } 
        
        \put( 0.95, 11.9){ \scalebox{1.1}{\makebox{AFM2}} }
        \put( 4.95, 11.9){ \scalebox{1.1}{\makebox{AFM1}} }
        \put( 0.95,  6.0 ){ \scalebox{1.1}{\makebox{AFM1}} }
        \put( 4.95,  6.0 ){ \scalebox{1.1}{\makebox{AFM1}} }
        
        \put(0.0, 9.2){
            \put( 0.85, 1.20){ \scalebox{1.0}{\makebox{Si}} }
            \put( 2.00, 0.30){ \scalebox{1.0}{\makebox{Mn2}} }
            \put( 1.20, 1.85){ \scalebox{1.0}{\makebox{Mn1}} } 
        }
        
        \put(3.75, 0.07){
            {\color{blue}
            \put( 1.82, 5.07){ \scalebox{1.0}{\makebox{$J_5$}} }
            \put( 1.15, 4.45){ \scalebox{1.0}{\makebox{$J_6$}} }
            \put( 0.62, 4.32){ \scalebox{1.0}{\makebox{$J_4$}} }
            \put( 1.47, 3.90){ \scalebox{1.0}{\makebox{$J_2$}} }
            \put( 1.37, 3.35){ \scalebox{1.0}{\makebox{$J_1$}} }
            \put( 1.58, 2.92){ \scalebox{1.0}{\makebox{$J_3$}} }
            }
        }
        
        \put(-0.7,0.0) {
            \put( 0.05, 0 ){ \includegraphics[width=1.7cm,trim={30cm 29cm 10cm  10cm  },clip=true]{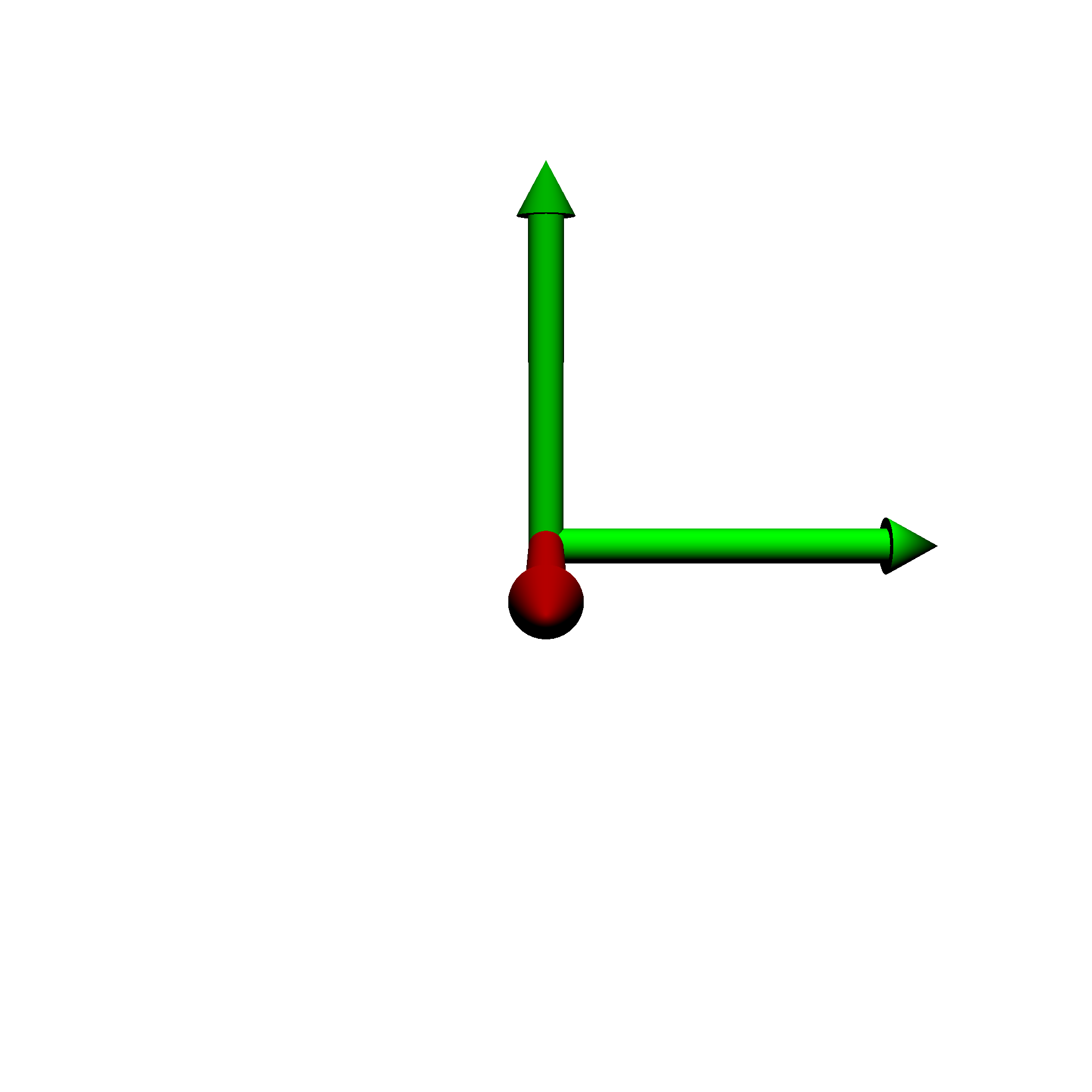} }
            \put( 1.9, 0.1){ \scalebox{0.9}{\makebox{a}} }
            \put(   0, 1.9){ \scalebox{0.9}{\makebox{b}} }
            \put(   0, 0.1){ \scalebox{0.9}{\makebox{c}} }
        }
        
        \put(5.16,2.73){
        \scalebox{0.72}{\linethickness{0.7pt}\polygon(0,0)(0.5,-1.4)(1,0)}
        }
    }
\end{picture}
\caption{
Magnetic structures of $\textnormal{Mn}_5\textnormal{Si}_3$.
  (a) The collinear AFM2 phase according to Ref.~\onlinecite{brown_antiferromagnetism_1995,gottschilch_study_2012,dos_Santos_2021}. 
  Proposed noncollinear AFM1 phase from (b) Ref.~\onlinecite{brown_low-temperature_1992,comment2}, (c) Ref.~\onlinecite{gottschilch_study_2012} and (d) the present work.
  Sites occupied by Mn1, Mn2 and Si atoms are shown with red, magenta and yellow spheres, respectively. The green arrows depict the orientation of the magnetic moments and the blue solid lines indicate the relevant exchange interactions used in the Heisenberg Hamiltonian (see details in text). The black triangle highlights a noncollinear spin arrangement formed by the Mn1 and Mn2 magnetic moments, which is a common feature to all proposed AFM1 magnetic structures.
}
\label{fig:all_AFM1}
\end{figure}

In the paramagnetic (PM) state, $\textnormal{Mn}_5\textnormal{Si}_3$ crystallizes in the hexagonal space group $P6_{3}/mcm$ with two distinct crystallographic positions for the Mn atoms (sites Mn1 and Mn2)~\cite{gottschilch_study_2012}. 
With decreasing temperature the onset of AFM orders (first at $T_{N_{2}}\approx$ 100\,K and then at $T_{N_{1}}\approx$ 60\,K) results in a reduction of the crystal symmetry. For temperatures between $60 < T < 100$\,K (AFM2 phase), the crystal structure can be described by a centrosymmetric orthorhombic cell with space group $Ccmm$, where Mn2 divides into two sets of nonequivalent positions~\cite{brown_antiferromagnetism_1995,gottschilch_study_2012}. In this cell, magnetic reflections follow the condition $h+k$ odd, the magnetic propagation vector is $\bm\kappa = (0, 1, 0)$, and only two-thirds of the Mn2 atoms acquire magnetic moments aligned parallel and antiparallel to the $b$ axis of the orthorhombic unit cell~\cite{brown_antiferromagnetism_1995,gottschilch_study_2012}(see Fig.~\ref{fig:all_AFM1}(a) with a more detailed discussion in the upcoming sections). In addition, recently performed DFT calculations are in line with the experimentally established collinear magnetic structure of the AFM2 phase~\cite{dos_Santos_2021}.
For $T < 60$\,K (AFM1 phase) the crystal symmetry is further reduced, the magnetic moments reorient in a highly noncollinear and noncoplanar arrangement, while the propagation vector remains the same as in the AFM2 phase. Albeit the magnetic structure has monoclinic~\cite{gottschilch_study_2012,brown_low-temperature_1992,silva_magnetic_2002} or possibly lower symmetry~\cite{comment}, the atomic positions can be described with an orthorhombic cell without inversion symmetry (space group $Cc2m$)~\cite{brown_low-temperature_1992,silva_magnetic_2002}. 
According to the proposed magnetic structures (see Figs.~\ref{fig:all_AFM1}(b) and \ref{fig:all_AFM1}(c)), the AFM1 phase is quite complex and rather unusual, as the Mn atoms acquire different magnetic moments even if they have similar chemical environments.
Despite the controversy regarding the spin orientation in the AFM1 phase, it is accepted that not only two-thirds of the Mn2 (as in the AFM2 phase) but also the Mn1 atoms carry a magnetic moment leaving still one-third of the Mn2 atoms without moment~\cite{gottschilch_study_2012,Lander_1967,brown_low-temperature_1992}. 

\begin{figure}[tb]
\centering
\includegraphics[width=8.5 cm,trim={1cm 0 0cm 0},clip=true]
{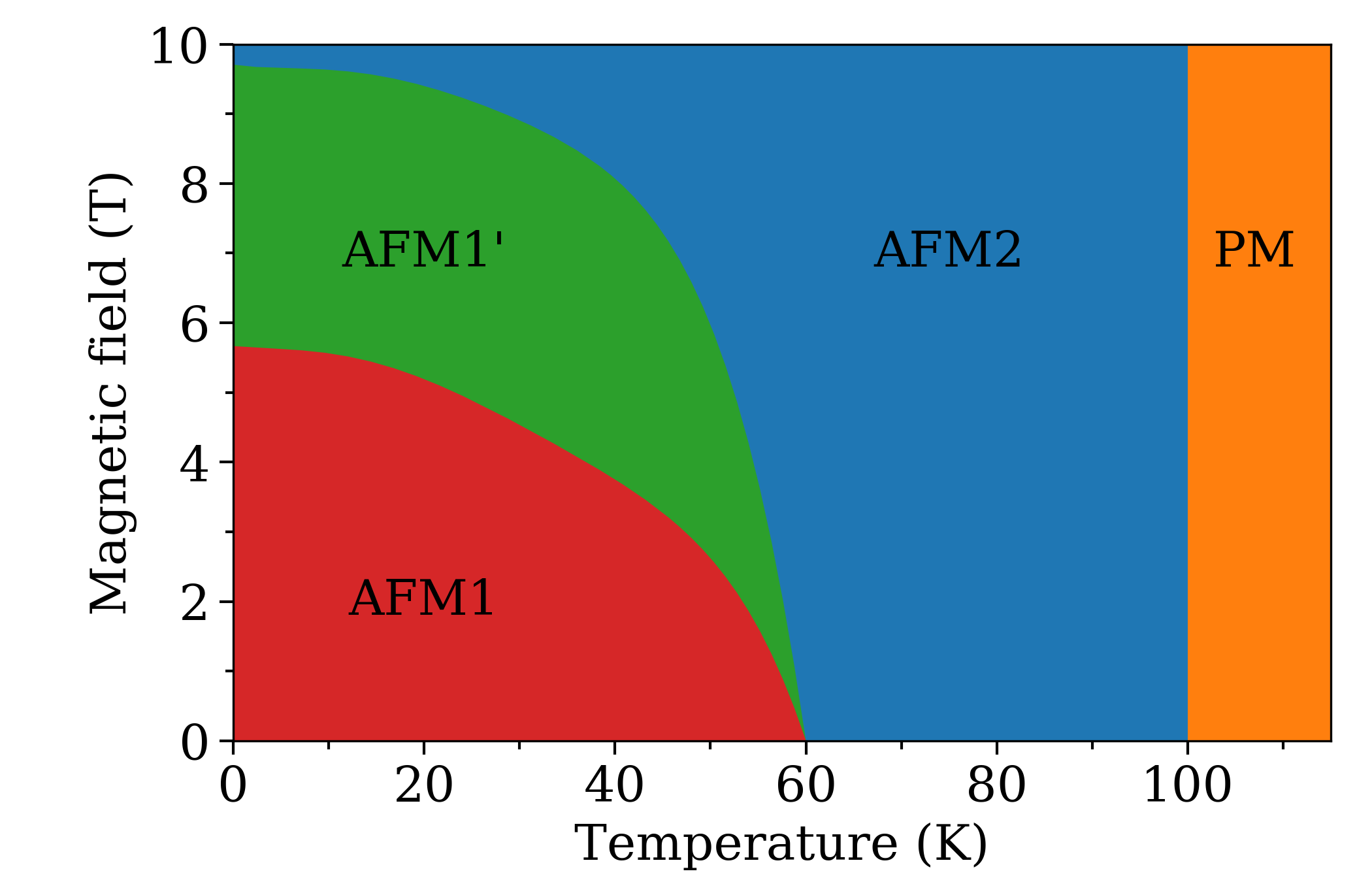}
\caption{
Temperature and magnetic field phase diagram of $\textnormal{Mn}_5\textnormal{Si}_3$ for $\VEC B\parallel \hat{\VEC c}$ based on Ref.~\onlinecite{surgers_switching_2017}.
The borders between the phases are determined from different macroscopic measurements: magnetization, resistivity, and Hall effect.
}
\label{fig:phase_diagram}
\end{figure}

In $\textnormal{Mn}_5\textnormal{Si}_3$, apart form the reduction of temperature, the application of an external magnetic field results in field-induced transitions~\cite{surgers_switching_2017,Suergers_2016,das_observation_2019,Alkanani_1995}. The magnetic phase diagram as a function of temperature and magnetic field as established by magnetization and electrical transport measurements is shown in Fig.~\ref{fig:phase_diagram}.
A very steep phase boundary $T_{N_{2}}(B)$ is outlined between the AFM2 phase and the PM state and in the temperature range where the AFM2 phase is observed, no field-induced transition is reported up to the maximum investigated field of 10\,T.
Below 60\,K, the increasing magnetic field precipitates transitions from the AFM1 phase to another intermediate AFM phase~\cite{surgers_switching_2017,das_observation_2019} (marked as AFM1' in Fig.~\ref{fig:phase_diagram}) before reaching the collinear AFM2 phase. Neutron scattering investigations under field in single crystal~\cite{biniskos_spin_2018,silva_magnetic_2002} and polycrystalline~\cite{gottschilch_study_2012} samples confirm the existence of these phase transitions.
We also note that for $\textnormal{Mn}_5\textnormal{Si}_3$ a modified $B-T$ phase diagram from the one shown in Fig.~\ref{fig:phase_diagram} was proposed for flux grown single crystals~\cite{luccas_magnetic_2019}. The differences were suggested to originate from the inherent stress that the samples acquire when grown by different techniques. 

\section{Methods}

\subsection{Experimental details}

Two single crystals (with mass of about 7\,g each) of $\textnormal{Mn}_5\textnormal{Si}_3$ were individually mounted on an aluminium sample holder and oriented in the [100]/[010] and [010]/[001] scattering plane of the orthorhombic symmetry, respectively. The single crystals were grown by the Czochralski method and are the same that were used in previous studies~\cite{biniskos_spin_2018,dos_Santos_2021}. The spin waves in the AFM1 phase of $\textnormal{Mn}_5\textnormal{Si}_3$ were investigated as a function of the wave-vector $\VEC Q$ and the energy transfer $E$ at $T = 10$\,K. In this work we use the orthorhombic coordinate system and the scattering vector is expressed in Cartesian coordinates $\VEC Q = (Q_{h}, Q_{k}, Q_{l})$ given in reciprocal lattice units (r.l.u.). The wave-vector $\VEC q$ is related to the momentum transfer through $\hbar\VEC Q = \hbar\VEC G + \hbar\VEC q$, where $\VEC G$ is an AFM zone center and $\VEC G = (h, k, l)$.

INS experiments were carried out at the Institut Laue-Langevin (ILL) using the cold and thermal neutron three-axis spectrometers (TASs) IN12~\cite{schmalzl_upgrade_2016} and IN22, respectively. The instrument resolution was in each case adapted to the studied momentum and energy range. Both TASs were setup in W configuration and inelastic scans were performed with constant $\VEC{k}_f$, where $\VEC{k}_f$ is the wave-vector of the scattered neutron beam. A pyrolytic graphite (PG(002)) monochromator and analyzer were used. Higher-order contamination was removed using a velocity selector (at IN12) and a PG filter (at IN22) before the monochromator and in the scattered neutron beam, respectively. The single crystals were cooled below room temperature with a $^4$He flow cryostat. The neutron data collected at ILL are available at \onlinecite{data_IN12a,data_IN12b,data_IN22}.

\subsection{Modeling the AFM1 phase of $\textnormal{Mn}_5\textnormal{Si}_3$}

The starting point for modeling the noncollinear phase (AFM1) of $\textnormal{Mn}_5\textnormal{Si}_3$ is the Heisenberg Hamiltonian proposed in a recent study in Ref.~\onlinecite{dos_Santos_2021}, where first-principles calculations were performed to determine the ground state electronic and magnetic properties of the collinear phase (AFM2).
In the following paragraph, the methods and the relevant results of Ref.~\onlinecite{dos_Santos_2021} are briefly summarized.

Density functional theory was employed using the full-potential Korringa-Kohn-Rostoker Green-function (KKR-GF) method including spin-orbit coupling, as implemented in the \textit{JuKKR} code~\cite{papanikolaou_conceptual_2002}, using the local spin density approximation~\cite{vosko_accurate_1980}.
The magnetic exchange tensor, which parametrizes the spin Hamiltonian, was obtained through the infinitesimal rotations method~\cite{liechtenstein_local_1987,ebert_anisotropic_2009}. The Hamiltonian reads as:

\begin{equation}\label{eq:heisenberg_hamiltonian}
    \HH = - \sum_{ij} J_{ij} \VEC S_i \cdot \VEC S_j  - \sum_\alpha k^\alpha \sum_i (S^\alpha_i)^2 ,
\end{equation}
where the first term captures the magnetic exchange interactions and the second term accounts for the biaxial magnetocrystalline anisotropy.
$\VEC S_i$ refers to the spin, which is set to $S = 1$. 
The magnetic exchange interactions for the AFM2 phase were obtained from first-principles calculations.
The DFT calculations also indicated that $b$ and $c$ are the primary and the secondary easy-axis, respectively~\cite{dos_Santos_2021}, with $k^b = 0.12$\,meV per magnetic atom (meV/p.m.a) and $k^c = 0.03$\,meV/p.m.a.
To match the INS data for the spin-wave gap at the magnetic zone center, the authors in Ref.~\onlinecite{dos_Santos_2021} set $k^c = 0.09$\,meV/p.m.a. and scaled down uniformly the DFT parameters, exchange interactions and anisotropy constants, by a factor of 10.

\begin{table}[h]
\begin{ruledtabular}
\begin{tabular}{ccdd}
Parameter & Type & \multicolumn{1}{c}{Value (meV)} & \multicolumn{1}{c}{Distance (\AA)} \\
\hline
$J_1$ & Mn2--Mn2 & -12.23 & 2.825 \\
$J_2$ & Mn2--Mn2 & -2.16 & 2.907 \\
$J_3$ & Mn2--Mn2 & +3.98 & 4.054 \\
$J_4$ & Mn2--Mn2 & -2.89 & 4.371 \\
$J_5$ & Mn1--Mn1 & +11.99 & 2.407 \\
$J_6$ & Mn1--Mn2 & -2.17 & 2.959
\end{tabular}
\end{ruledtabular}
\caption{\label{tab:exchange_interactions}
Calculated exchange constants $J_{ij}$ for $\textnormal{Mn}_5\textnormal{Si}_3$ that stabilize our proposed magnetic structure for the noncollinear AFM1 phase. The distance refers to the corresponding Mn--Mn bond length. Positive (negative) values characterize FM (AFM) coupling.
}
\end{table}

Based on refinements on neutron diffraction data~\cite{gottschilch_study_2012,Lander_1967,brown_low-temperature_1992}, the Mn1 sites of the noncollinear phase (AFM1) of $\textnormal{Mn}_5\textnormal{Si}_3$ acquire a finite magnetic moment.
As we are technically limited to extract exchange interactions from collinear phases, we performed a DFT calculation with the magnetic moments in a ferromagnetic (FM) configuration~\cite{dos_Santos_2021}.
In this state, the Mn1 sites have a finite magnetic moment (about half of the moment in the Mn2 sites, similar to the value obtained in Ref.~\onlinecite{brown_low-temperature_1992}), which then allows us to calculate its exchange interactions with the other moments.
Using this calculation, one can estimate the exchange coupling between the nearest-neighbors Mn1--Mn1 ($J_{5}$) and Mn1--Mn2 ($J_{6}$) interactions.
Thus, in the model introduced in the previous paragraph, we add a finite magnetic moment in the Mn1 sites, a FM coupling between the Mn1 sites along the $c$ axis, and an AFM coupling between the Mn1 and Mn2 sites (see Table~\ref{tab:exchange_interactions}).

In order to investigate the AFM1 phase of $\textnormal{Mn}_5\textnormal{Si}_3$, we consider an orthorhombic cell as described in Ref.~\onlinecite{brown_low-temperature_1992} and we used the Hamiltonian parameters as obtained from DFT without any rescaling or adjustment.
The values for the magnetic exchange interactions are shown in Table~\ref{tab:exchange_interactions} and the parameters for the biaxial magnetic anisotropy are $k^b = 0.12$\,meV/p.m.a and $k^c = 0.03$\,meV/p.m.a.
As the INS measurements were performed at a low temperature ($T= 10$\,K), we do not need to rescale the Hamiltonian parameters to account for thermal fluctuations when modelling the spin-wave spectrum.
In Fig.~\ref{fig:all_AFM1}d and in Table~\ref{tab:exchange_interactions}, $J_1$, $J_2$ and $J_3$ correspond to couplings between the Mn2 spins in the same $\textnormal{[Mn2]}_6$ octahedra, $J_4$ refers to the interaction between Mn2 spins located in adjacent $\textnormal{[Mn2]}_6$ octahedra, $J_5$ couples the Mn1--Mn1 spins along the $c$ direction and $J_6$ concerns the shortest distance between Mn1--Mn2 spins.

Spin dynamics simulations using the Spirit code~\cite{muller_spirit:_2019} were performed for determining the ground-state spin configuration (see Fig.~\ref{fig:all_AFM1}(d)).
After the completion of this step, the spin-wave excitations of the quantum Heisenberg Hamiltonian were obtained by employing the linear spin-wave approximation.
The spin-wave excitations are the eigenstates of the dynamical matrix associated with the quantum Heisenberg Hamiltonian in Eq.~\eqref{eq:heisenberg_hamiltonian}, as explained in detail in Ref.~\onlinecite{dos_santos_spin-resolved_2018}.
From the calculated magnon eigenvalues and eigenstates, the inelastic scattering spectra were derived using second-order time-dependent perturbation theory~\cite{dos_santos_spin-resolved_2018, dos_santos_nonreciprocity_2020, dos_santos_modeling_2020}.

\section{Results and Discussion}

\subsection{Spin waves in the AFM1 phase of $\textnormal{Mn}_5\textnormal{Si}_3$}

\begin{figure}[tb]
\setlength{\unitlength}{1cm} 
\newcommand{\boxsize}{0.3}
\begin{picture}(8,16.)
    \put(0.0, 10.8){\includegraphics[width=8 cm]{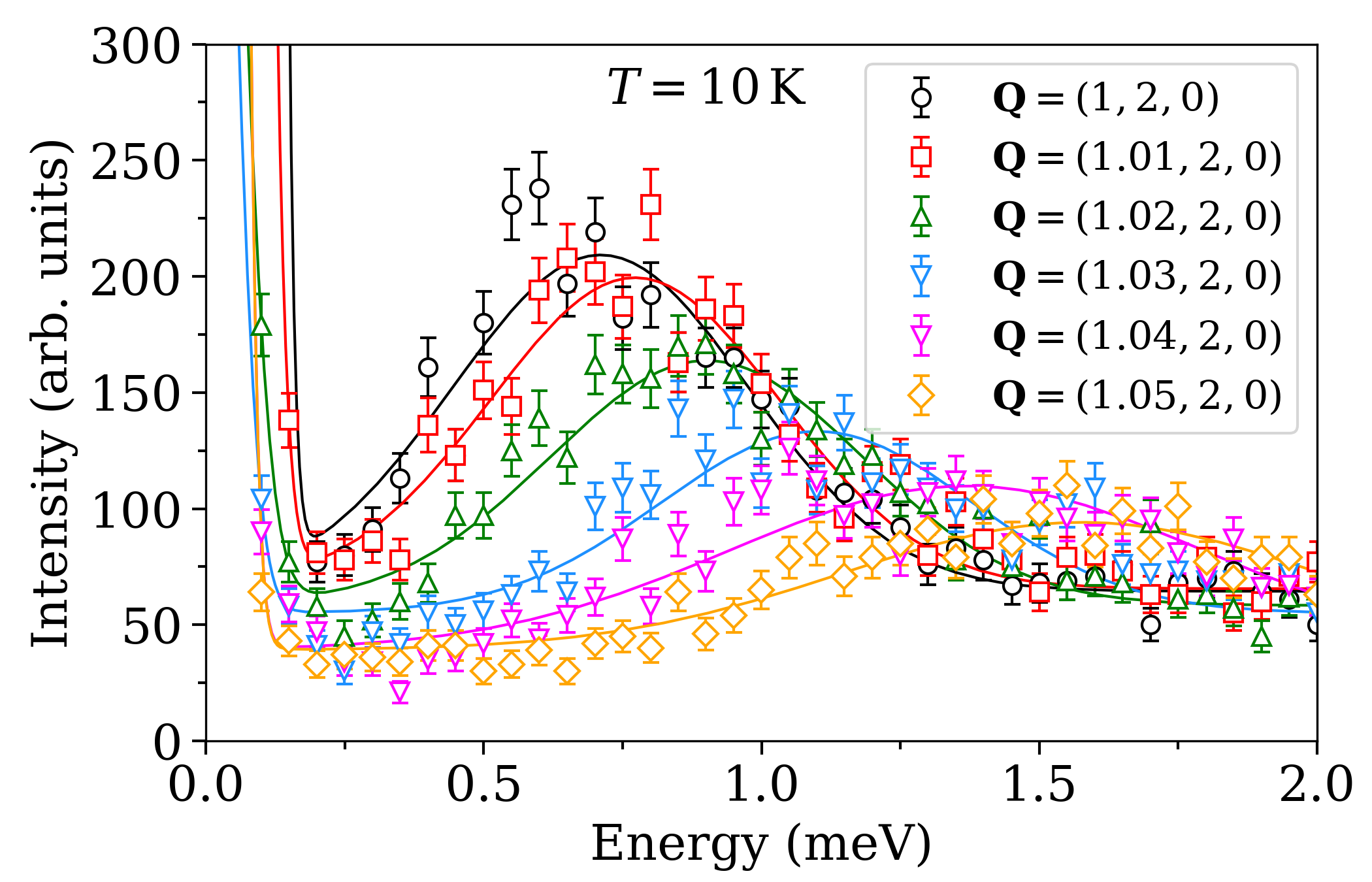}}
    \put(0.0, 5.4){\includegraphics[width=8 cm]{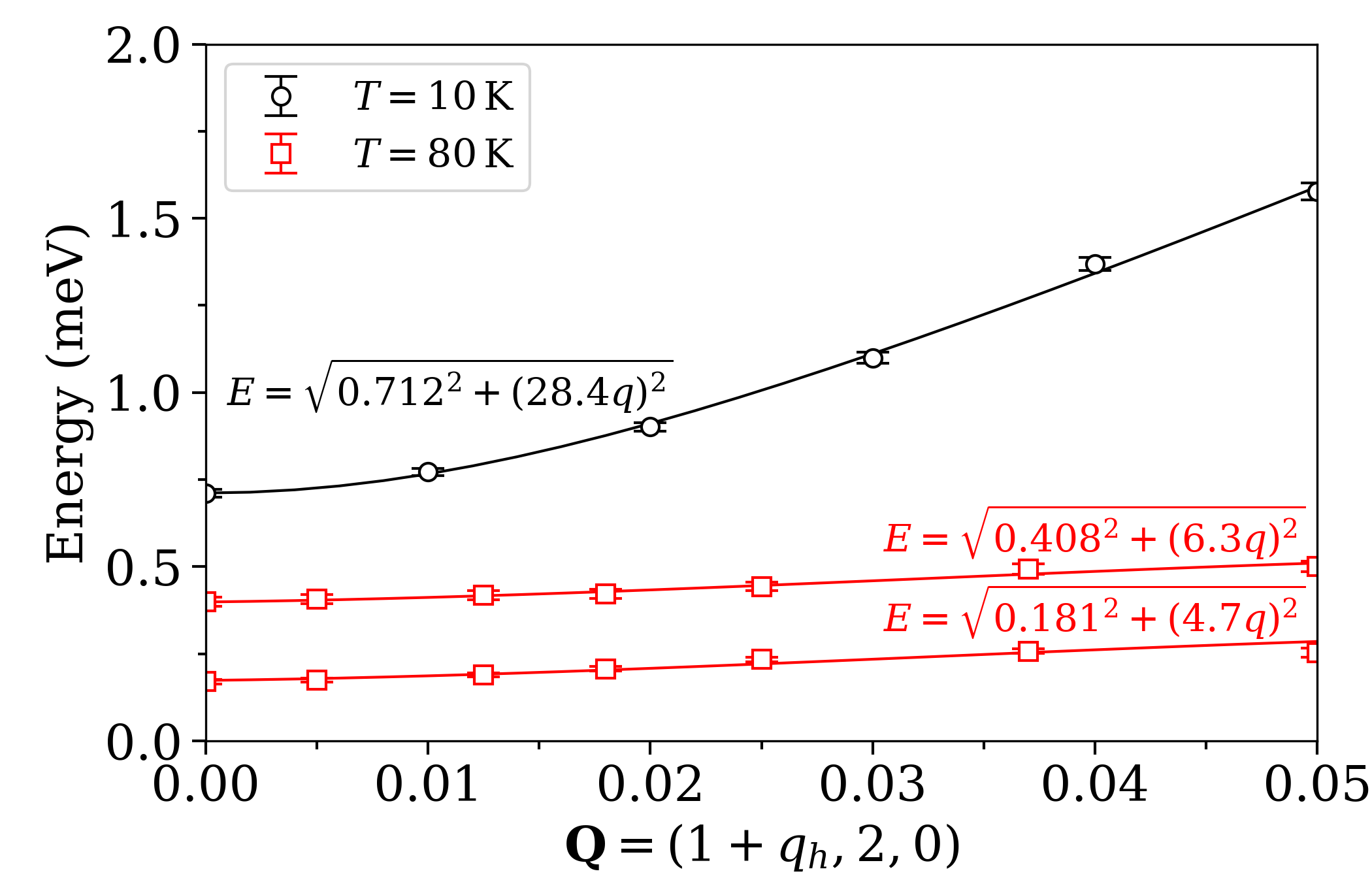}}
    \put(0.0, 0.0){\includegraphics[width=8 cm]{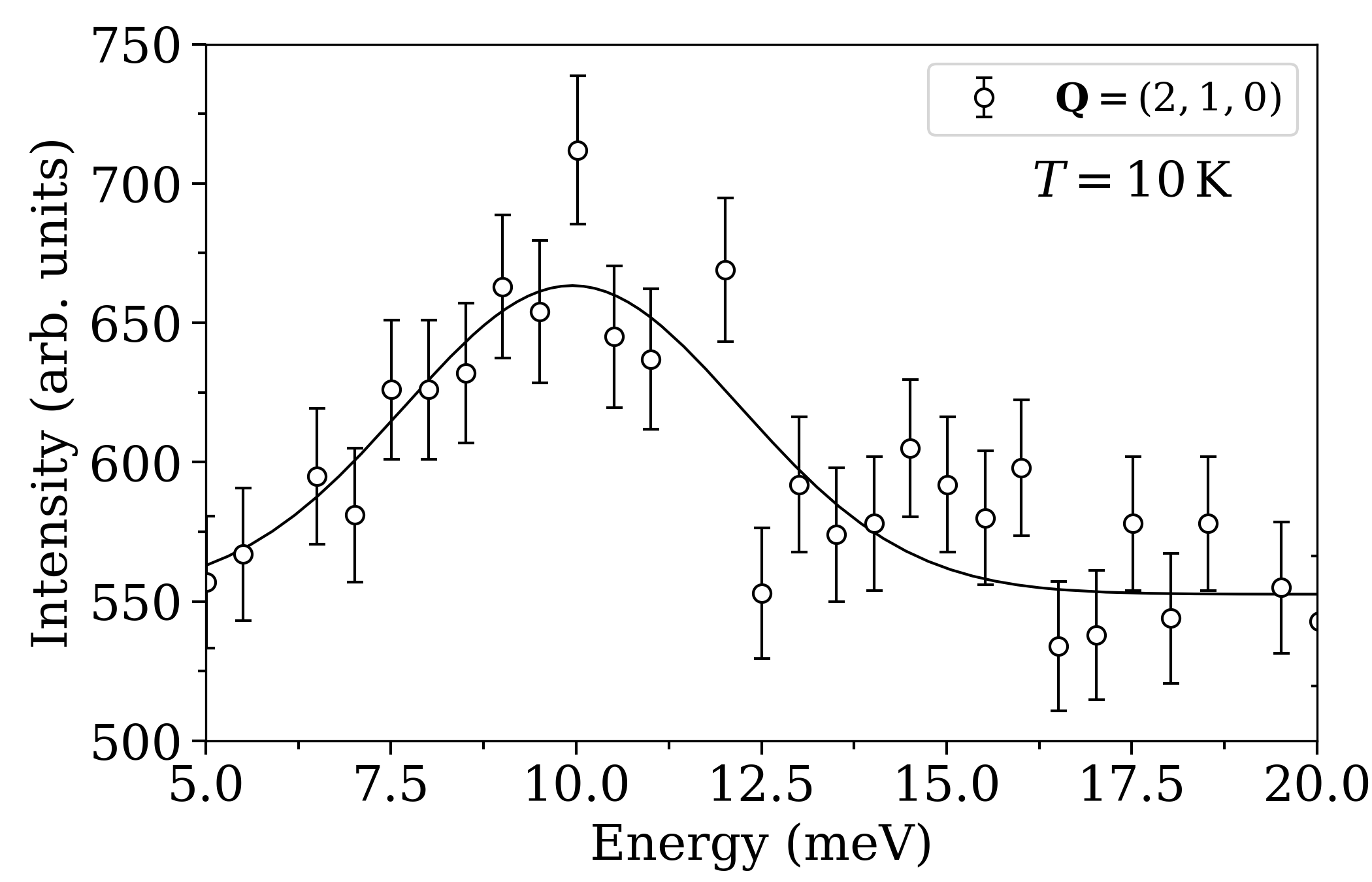}}
    
    \put(-0.4, 15.5){ \makebox(\boxsize,\boxsize){(a)} } 
    \put(-0.4, 10.1){ \makebox(\boxsize,\boxsize){(b)} } 
    \put(-0.4, 4.7){ \makebox(\boxsize,\boxsize){(c)} } 
\end{picture}
\caption{(a) Measured energy spectra at $\VEC Q = (Q_{h}, 2, 0)$ in the AFM1 phase of Mn$_{5}$Si$_{3}$. Data were obtained at $T = 10$\,K at IN12 with $k_f = 1.05$\,{\AA}$^{-1}$ and 40’-open-open collimations were installed. 
(b) Low energy magnon dispersion in the AFM1 (shown in black) and AFM2~\cite{dos_Santos_2021} (shown in red) phases along the ($h$00) direction.
(c) Energy scan at $\VEC Q = (2, 1, 0)$ at $T = 10$\,K obtained at IN22 with $k_f = 2.662$\,{\AA}$^{-1}$. 
The solid lines in (a) and (c) indicate fits with Gaussian functions, and in (b) fits with the empirical dispersion relation $E=\sqrt{\Delta^2+C^2q^2}$.}
\label{fig:spingap}
\end{figure}
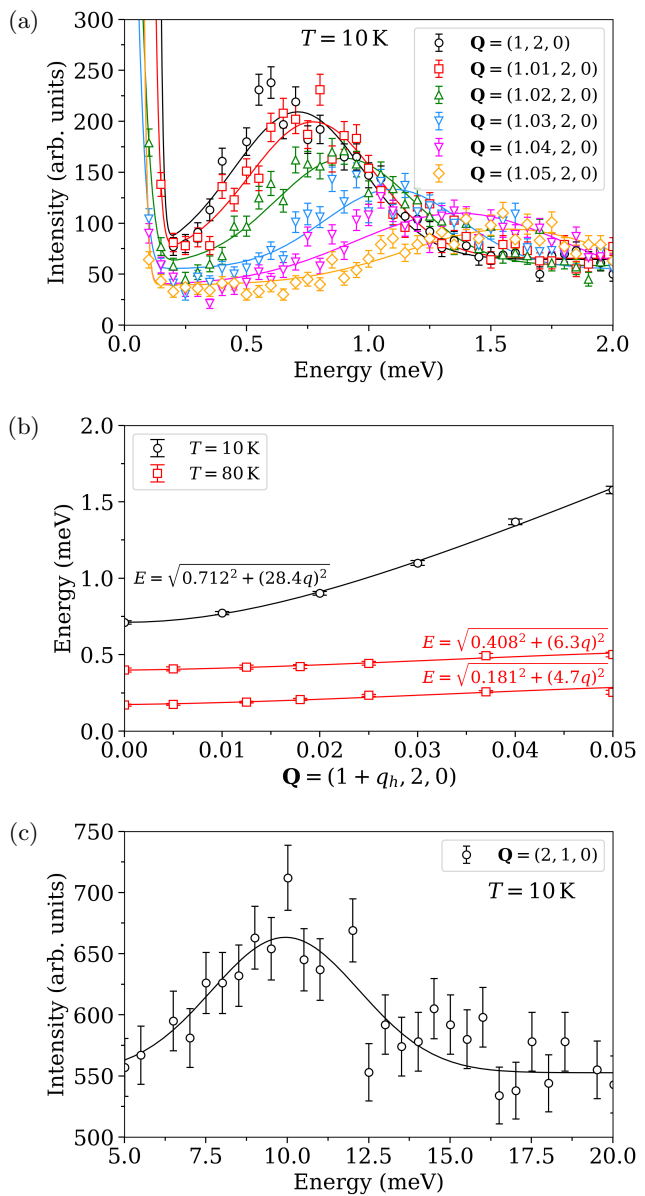

The investigation of the magnon spectrum in the noncollinear AFM1 phase was initiated with a collection of energy spectra around the magnetic zone center $\VEC G = (1, 2, 0)$ at $T = 10$\,K. Fig.~\ref{fig:spingap}(a) shows the energy dependence of the measured low energy excitations at different $Q_{h}$ positions, where $\VEC Q = (Q_h, 2, 0)$. In each $Q_{h}$ position the first peak is centered at $E = 0$\,meV and corresponds to the elastic line, while the second one at finite $E$, which propagates to higher energy transfers as $Q_{h}$ increases, unambiguously points to gapped spin waves. The low energy magnons along the ($h$00) direction in the AFM1 phase of Mn$_{5}$Si$_{3}$ are shown in Fig.~\ref{fig:spingap}(b) together with the dispersion of the AFM2 phase for comparison.

Similarly to the AFM2 phase, the data in the AFM1 phase can be described by the empirical dispersion relation $E=\sqrt{\Delta^2+C^2q^2}$~\cite{Ibuka} and the obtained values for the spin gap $\Delta$ and the constant $C_{(h00)}$ are 0.712(7)\,meV and 28.4(4)\,meV/r.l.u, respectively. Comparing the spin dynamics of two phases in the same ($\VEC q,E$) region reveals that the noncollinear AFM1 phase is characterized by a single gapped magnon branch at $\VEC q = 0$ in contrast to the collinear AFM2 phase where a splitting of the spin-wave modes is detected (double spin gap) due to the system’s biaxial anisotropy~\cite{dos_Santos_2021}. The spin gap at $T = 10$\,K (AFM1 phase) is about twice the gap at $T = 80$\,K (AFM2 phase) and its origin might be attributed to a local easy axis within this phase. One other important feature that becomes evident by comparing the values of the constants $C_{(h00)}$ is that the magnon dispersion in the AFM1 phase is about five times steeper than in the AFM2 phase. 

Further INS measurements with constant $\VEC Q$-scans were carried out around different magnetic zone centers for higher energy transfers where the kinematic constraints could be satisfied. Fig.~\ref{fig:spingap}(c) shows a spectrum measured at the magnetic zone center $\VEC G = (2, 1, 0)$ at $T = 10$\,K up to the energy transfer of 20\,meV. The observed peak can be described by a single Gaussian function and is assigned to an optical magnon branch that originates at about 10\,meV.

\begin{figure}[!tbh]
\setlength{\unitlength}{1cm} 
\newcommand{\boxsize}{0.3}
\begin{picture}(8,19.8)
    \put(0,-0.2){
        \put(0.0, 15.){\includegraphics[width=7.9 cm]{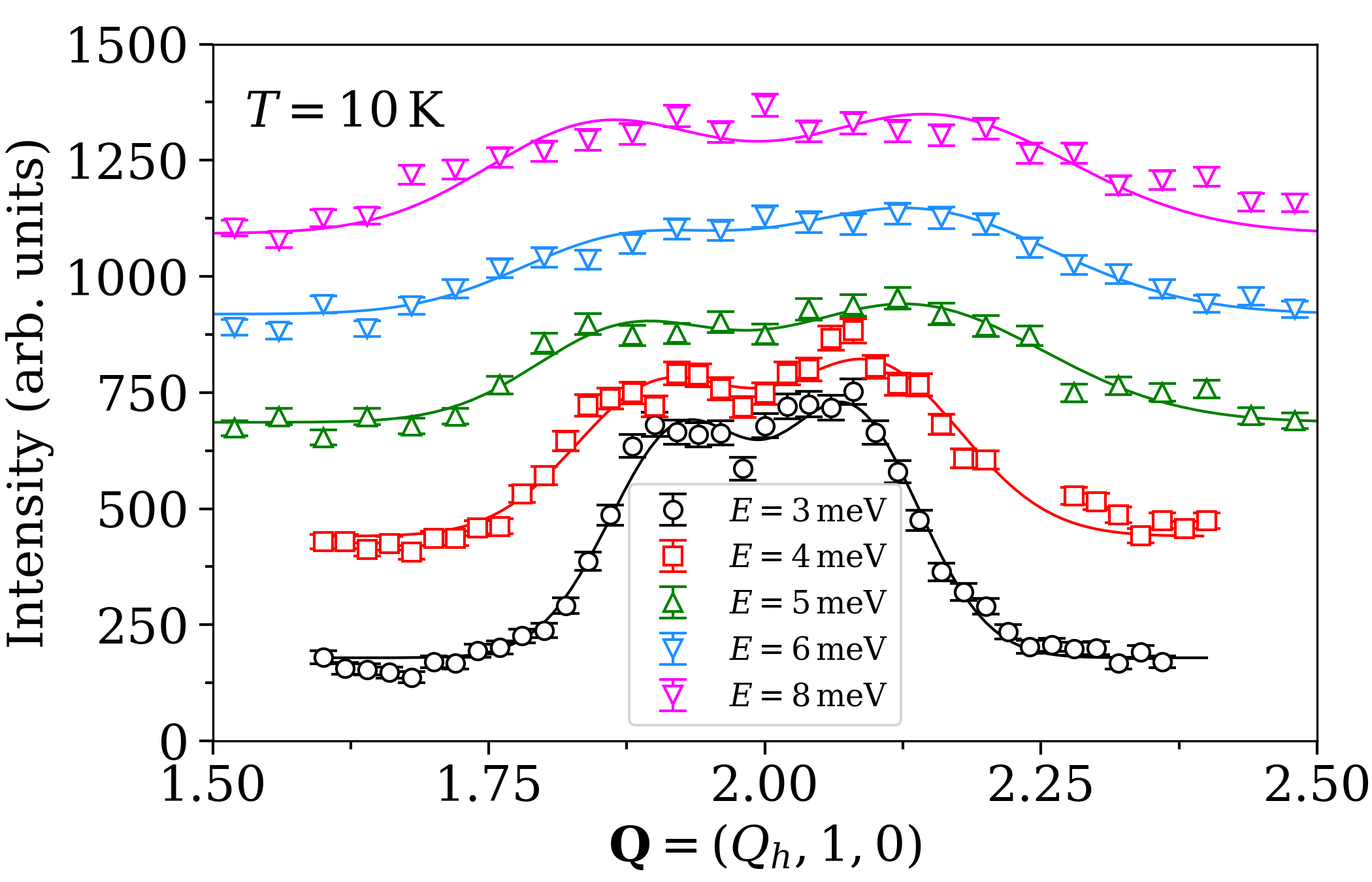}}
        \put(0.0, 10.){\includegraphics[width=7.9 cm]{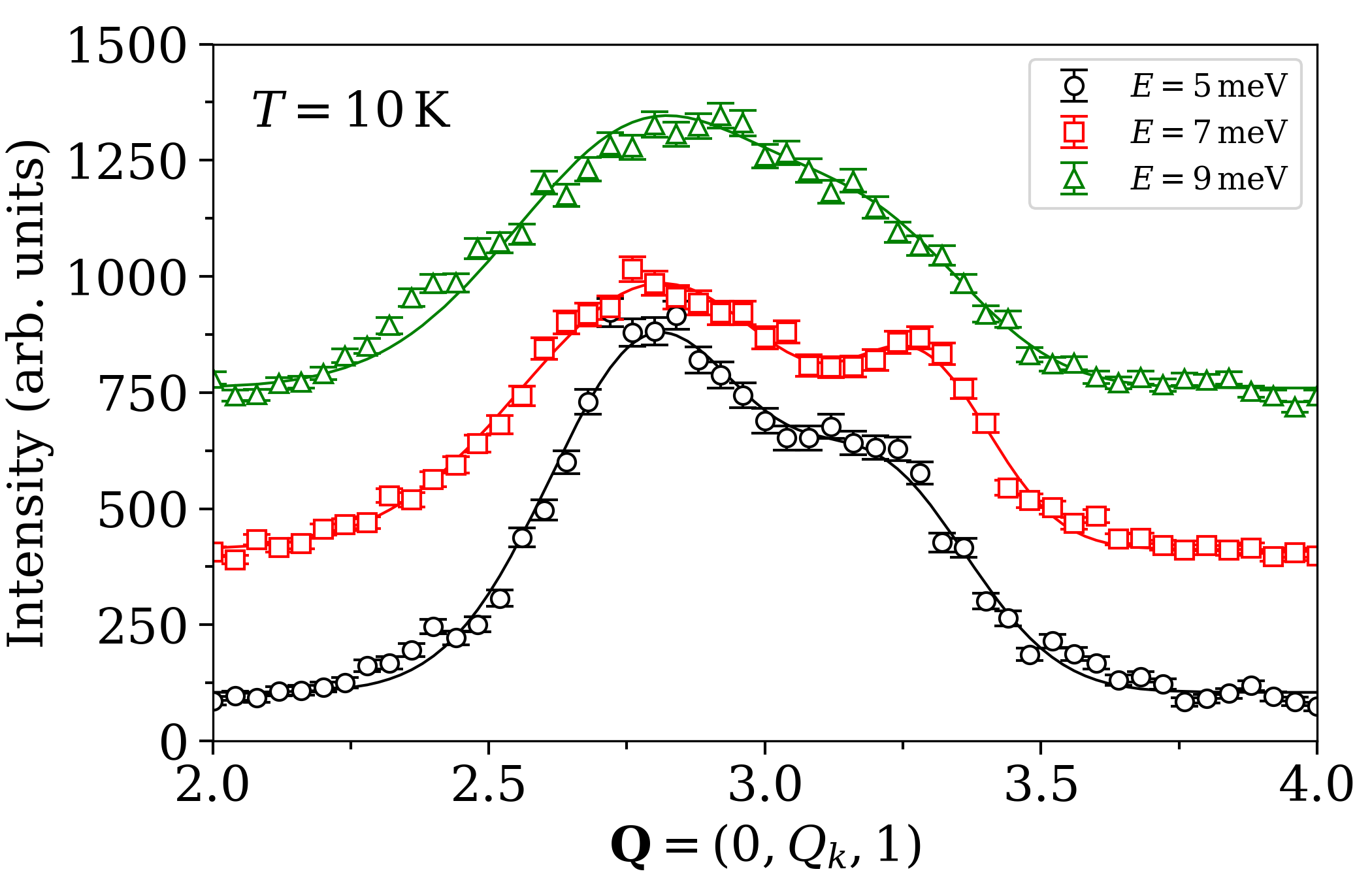}}
        \put(0.0, 5.){\includegraphics[width=7.9 cm]{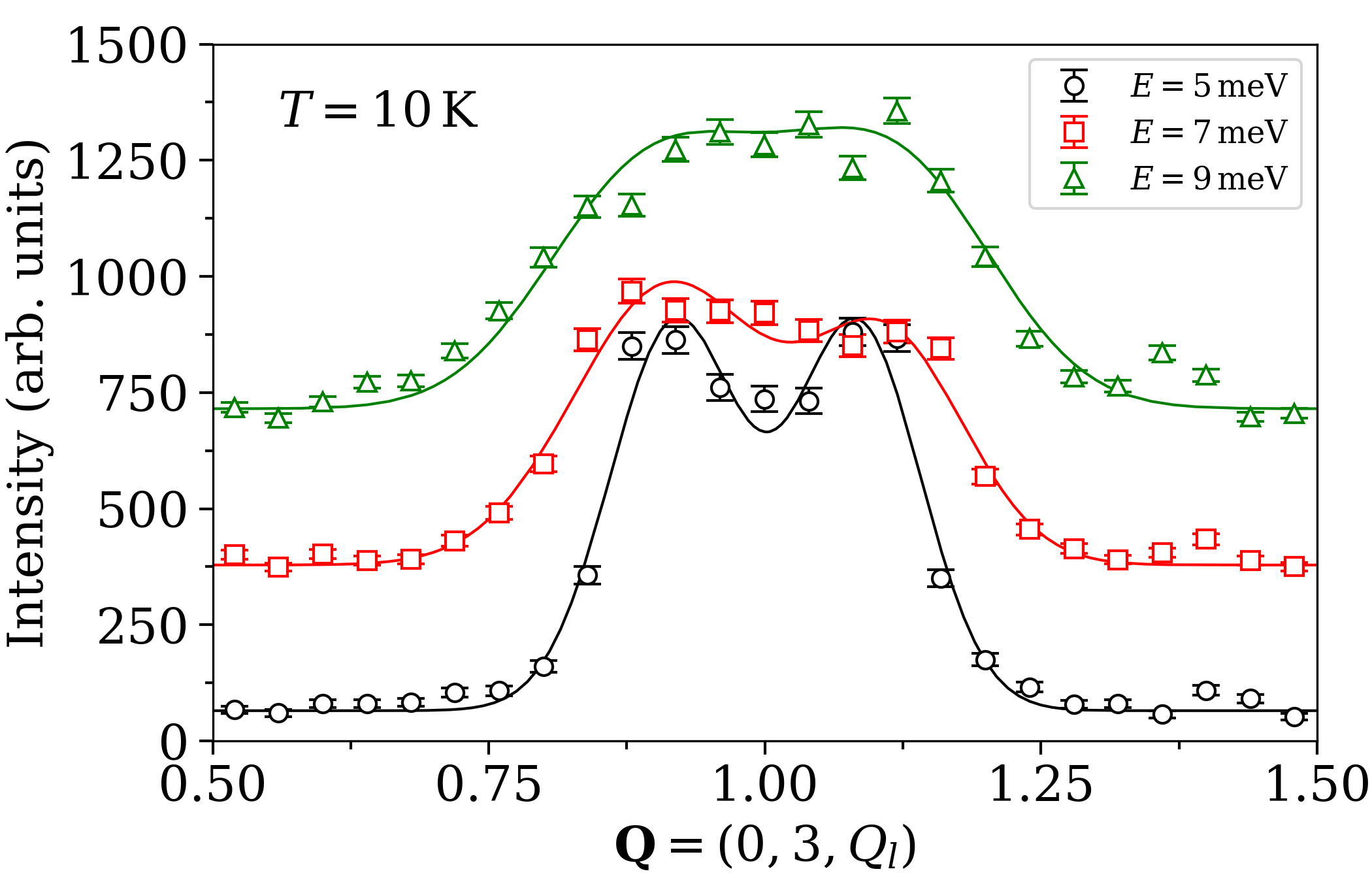}}
        \put(0.0, 0.0){\includegraphics[width=7.9 cm]{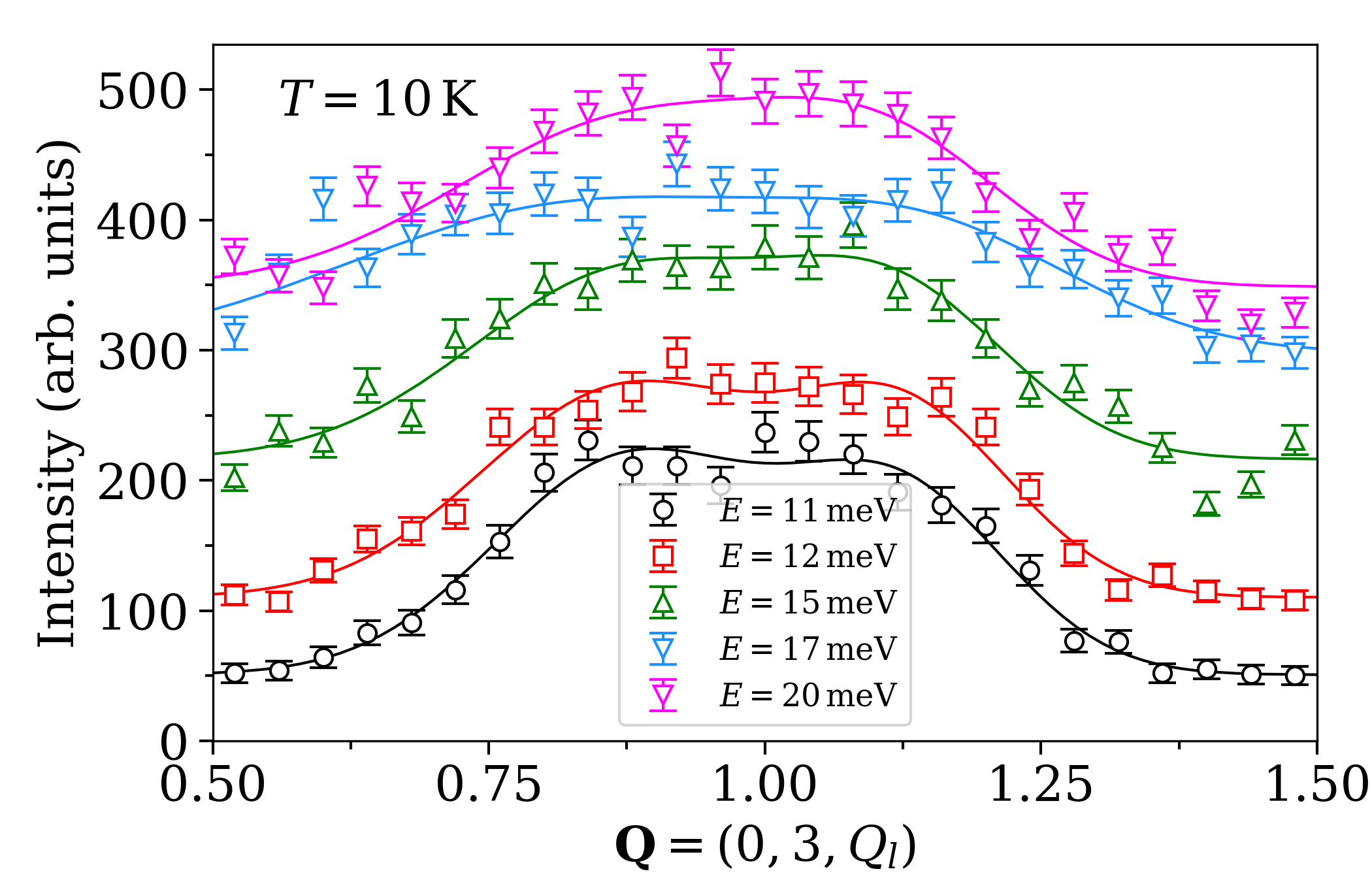}}

        \put(-0.5, 19.6){ \makebox(\boxsize,\boxsize){(a)} } 
        \put(-0.5, 14.6){ \makebox(\boxsize,\boxsize){(b)} } 
        \put(-0.5, 9.6){ \makebox(\boxsize,\boxsize){(c)} } 
        \put(-0.5, 4.6){ \makebox(\boxsize,\boxsize){(d)} } 
    }
\end{picture}
\caption{Inelastic spectra at different constant energy transfers measured at $T = 10$\,K at (a) $\VEC Q = (Q_{h}, 1, 0)$, (b) $\VEC Q = (0, Q_{k}, 1)$ and (c)-(d) $\VEC Q = (0, 3, Q_{l})$. The solid lines are fits with Gaussian functions. The spectra are shifted for clarity in intensity conserving the same scale. The data in (a), (d) and (b), (c) were obtained at IN22 (with $k_f = 2.662$\,{\AA}$^{-1}$) and IN12 (with $k_f = 2$\,{\AA}$^{-1}$), respectively.}
\label{fig:neutrondata}
\end{figure}
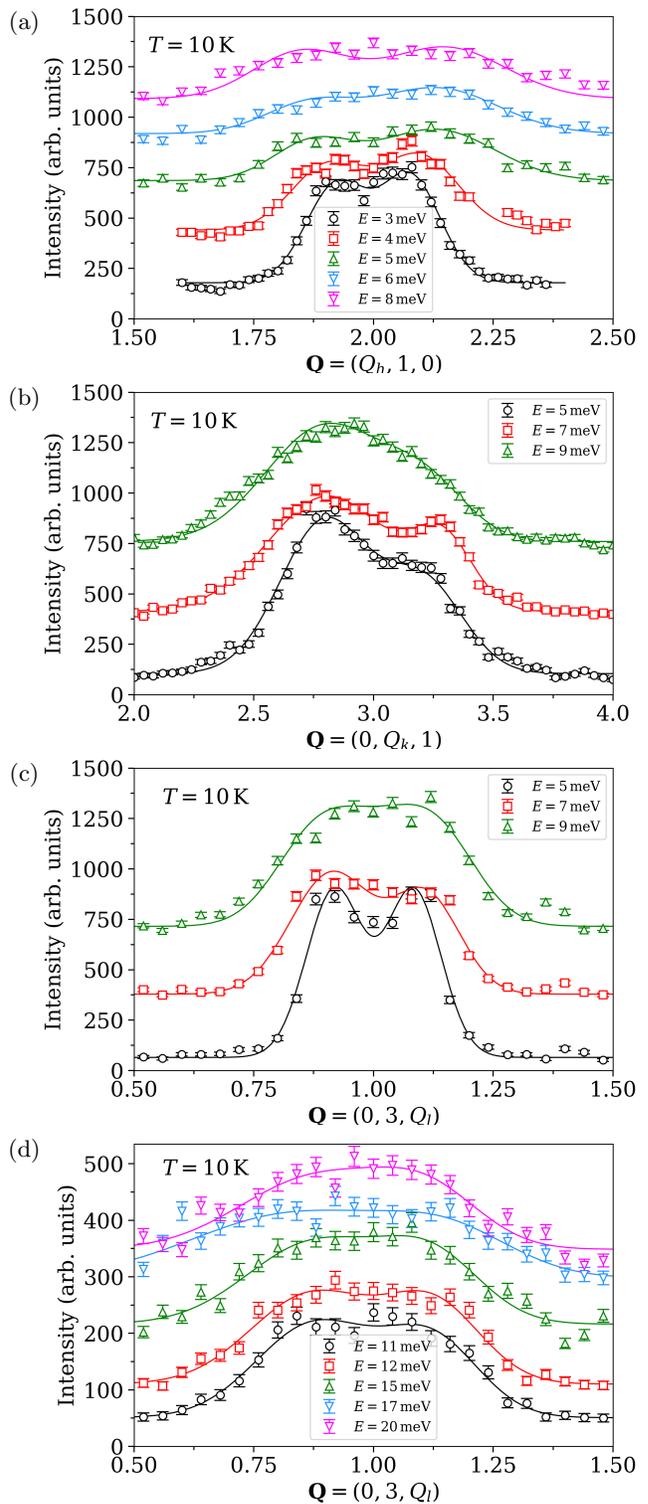

Since steep magnons were measured at low energies close to the magnetic zone center $\VEC G = (1, 2, 0)$, scans at constant $E$ were performed in order to obtain further the dispersion relations for higher energy transfers. Spin-wave excitations were measured along three high symmetry directions of the orthorhombic symmetry (($h$00), (0$k$0) and (00$l$)) around different AFM zone centers, namely $\VEC G = (2, 1, 0)$ and $\VEC G = (0, 3, 1)$. Fig.~\ref{fig:neutrondata} shows characteristic inelastic scans where the observed intensities that correspond to spin-wave scattering are fitted using double Gaussian functions on top of a flat background.
For $E < 10$\,meV, the peak widths increase with increasing energy transfer along all three directions (see Figs.~\ref{fig:neutrondata}(a)-(c)).
This broadening is independent of the instrumental resolution and is attributed to the contribution to the scattering intensities of an optical branch that is expected to originate at about 10\,meV (see Fig.~\ref{fig:spingap}(c)). For $E > 10$\,meV in the investigated $\VEC Q $ and energy range no further significant change in the peak widths is observed, as can be seen in the raw data shown in Fig.~\ref{fig:neutrondata}(d).
We note that the asymmetry of the scattering intensities which is observed in Fig.~\ref{fig:neutrondata}(b) leading to sharper and more intense spin-wave peaks at $-q$ compared to $+q$ can be attributed to the instrumental resolution focusing conditions.

\begin{figure}[tb]
\setlength{\unitlength}{1cm} 
\newcommand{\boxsize}{0.3}
  \begin{picture}(8.4,14)
    \put(0.0, 0.0){    
        \put(0., 0.0){ \includegraphics[width=8.5cm,trim={0 0.7cm 0 0},clip=true]{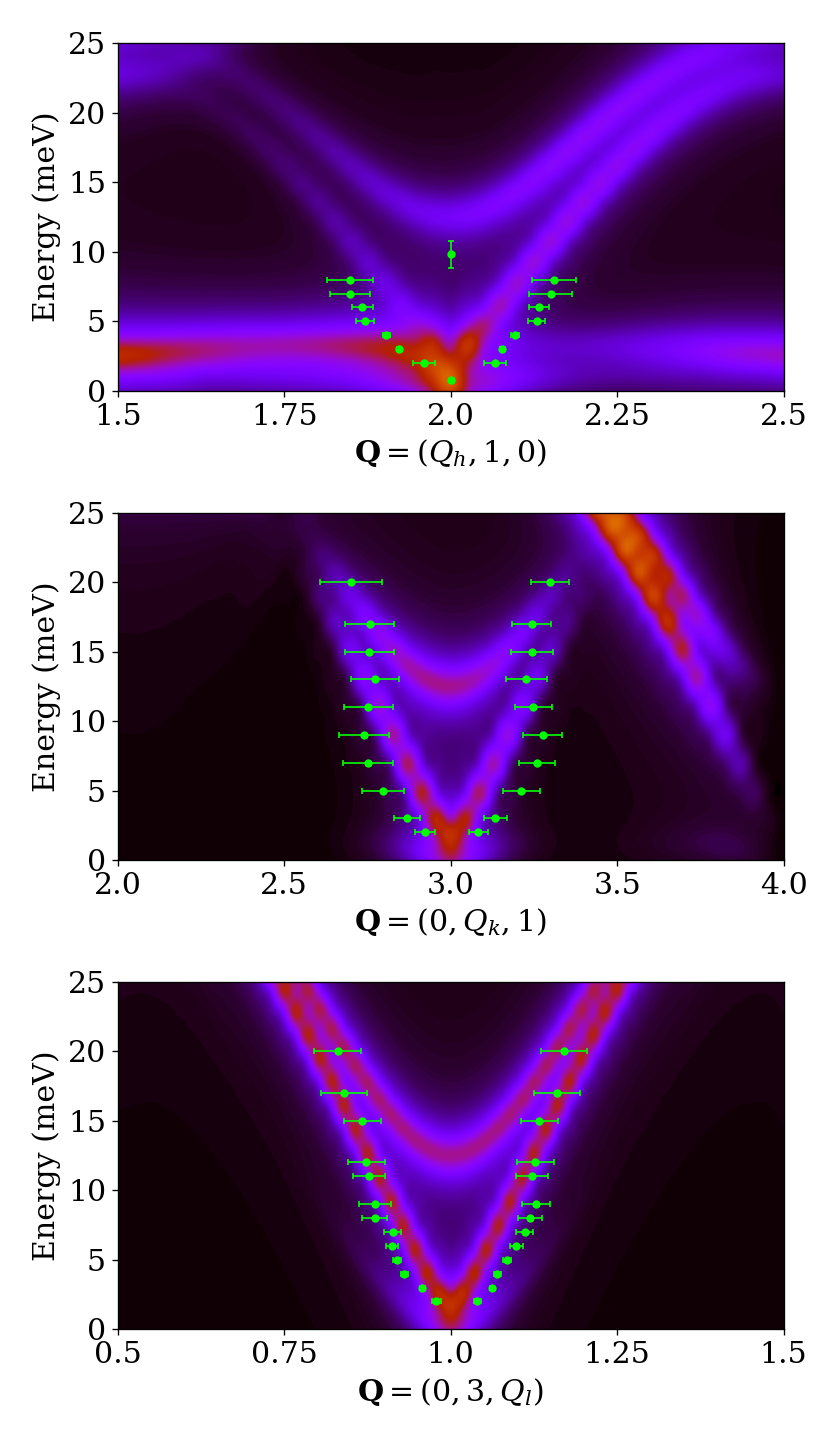}}
        
        \put(-0.2, 13.6){ \makebox(\boxsize,\boxsize){(a)} } 
        \put(-0.2, 8.83){ \makebox(\boxsize,\boxsize){(b)} } 
        \put(-0.2, 4.1){ \makebox(\boxsize,\boxsize){(c)} } 
    }
  \end{picture}

  \caption{\label{fig:theory_dispersions}
  Spin-wave dispersion relations of $\textnormal{Mn}_5\textnormal{Si}_3$ in the noncollinear AFM1 phase along the three high symmetry directions of the orthorhombic symmetry: (a) ($h$00), (b) (0$k$0) and (c) (00$l$). The data points are obtained from INS measurements and the color map corresponds to the calculated inelastic scattering signal.
  }
\end{figure}
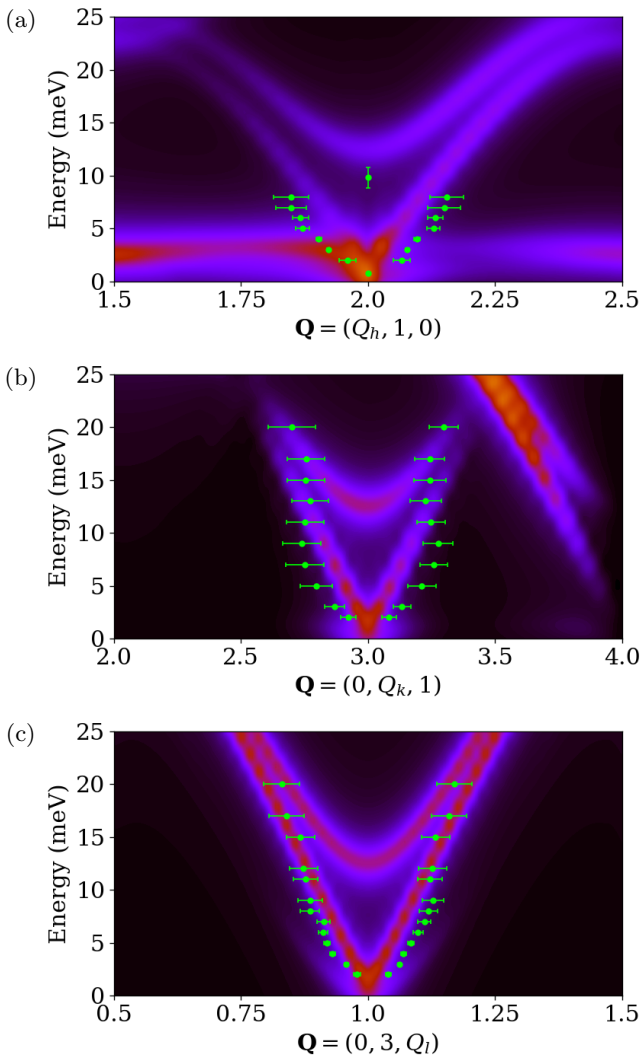

The experimentally and theoretically determined magnon dispersion relations of $\textnormal{Mn}_5\textnormal{Si}_3$ in the noncollinear AFM1 phase are shown in Fig.~\ref{fig:theory_dispersions}. 
The color mapping represents the intensity of the calculated inelastic scattering signal.
By using the minimal Hamiltonian (see Section II), we capture theoretically the main features of the magnons observed experimentally with INS along the three main crystal axes.
Usually, energy-dependent magnon damping is taken into account in theoretical calculations to reproduce some of the measured features of spin excitations in metallic magnetic systems~\cite{Diallo,Park_2018}.
In our INS data (see Figs.~\ref{fig:neutrondata}), we observe almost an uniform intensity throughout the whole measured energy range thus not requiring an energy-dependent broadening in the calculated spin-wave spectra.
Returning to the model results, the acoustic spin-wave modes are centered around the AFM zone centers and have the characteristic V-shape typically observed in several systems~\cite{Park_2018,Sukhanov_2019,Chen_2020,Harriger_2011,Wang_2015,Jacobsen_2018}.
An optic mode with an energy minimum of about 12\,meV is in good agreement with the one determined experimentally with INS (see Fig.~\ref{fig:spingap}(c)).
The height of this optic mode at the AFM $\Gamma$-point is strongly dependent on the $J_6$ exchange interaction, which couples the spins in the Mn1 and Mn2 sites.
In the theoretical scattering spectrum shown in Fig.~\ref{fig:theory_dispersions}(a), we observe also a flat feature at about 3\,meV.
According to our model this mode appears because of the existence of a magnetic moment in the Mn1 sites and its dispersionless behaviour is due to the small value of $J_6$.

\subsection{Field-induced phase transitions}

\begin{figure}[tb]
\setlength{\unitlength}{1cm} 
\newcommand{\boxsize}{0.3}

  \begin{picture}(8.4,11.5)
    
    \put(0.0, 0.2){    
        \put(0,0.4){
        \put(0.3, 7.1){
            \put(0.0, 0.0){\includegraphics[width=2.3 cm,trim={26.9cm 24.cm 26.9cm 24.cm},clip=true,frame]{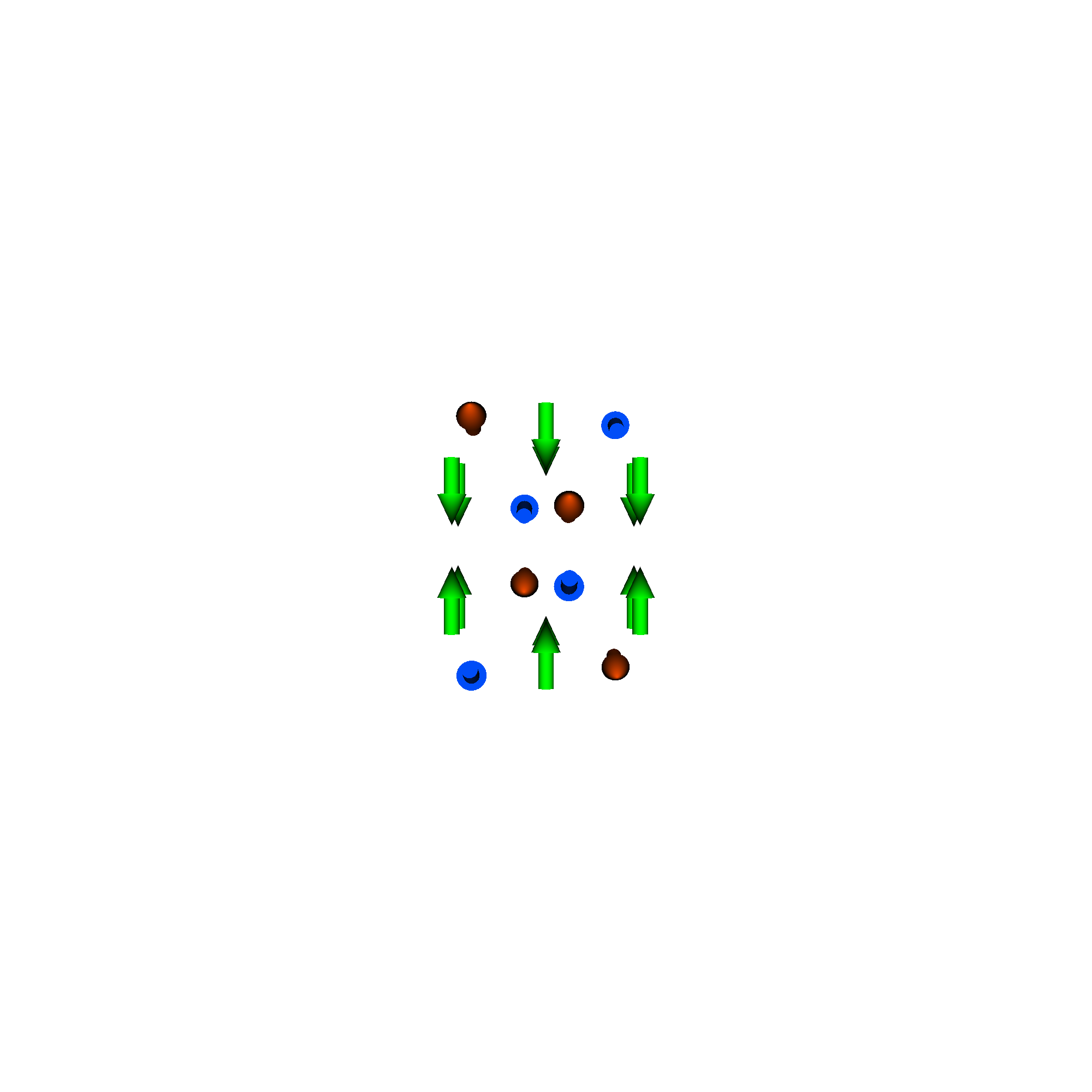}}
            \put(0.55, 3.00){ \scalebox{1.1}{\colorbox{white}{\makebox{0.0 T}}} }
            \put(1.75, 3.35){ \scalebox{1.1}{\makebox{AFM1}}}
        }
        \put(2.85, 7.1){
            \put(0.0, 0.0){\includegraphics[width=2.3 cm,trim={26.9cm 24.cm 26.9cm 24.cm},clip=true,frame]{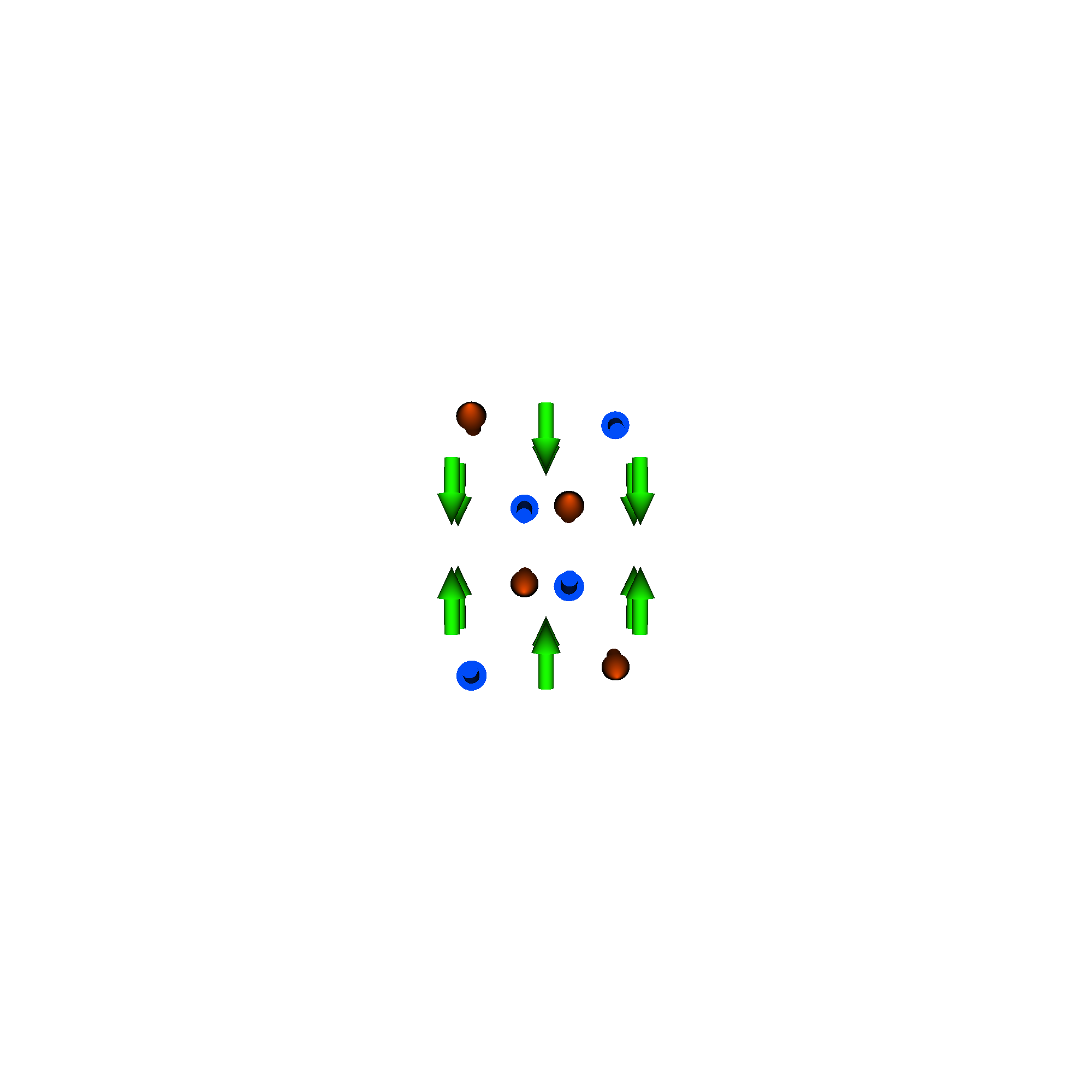}}
            \put(0.55, 3.00){ \scalebox{1.1}{\colorbox{white}{\makebox{2.3 T}}} }
        }
        \put(5.5, 7.1){
            \put(0.0, 0.0){\includegraphics[width=2.3 cm,trim={26.9cm 24.cm 26.9cm 24.cm},clip=true,frame]{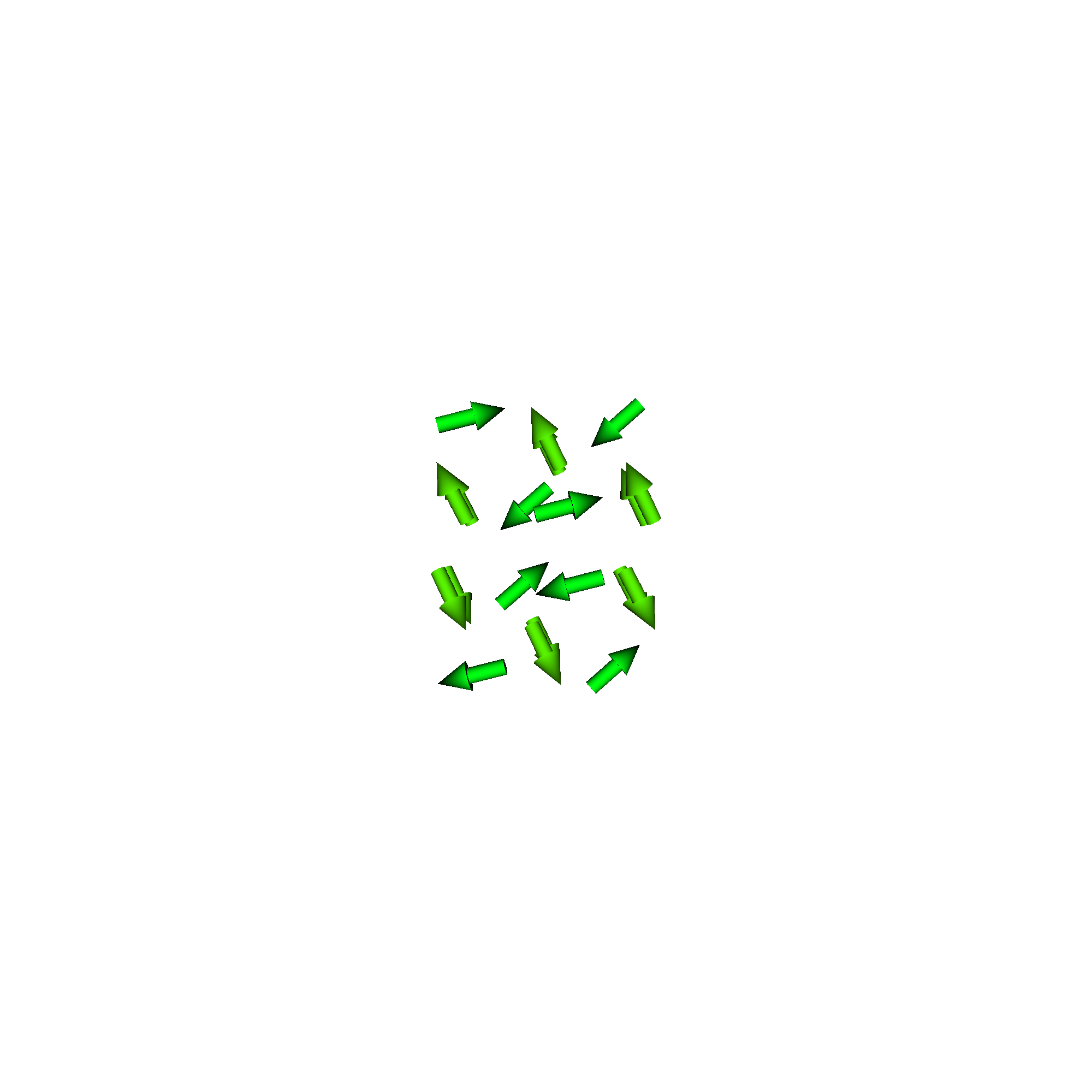}}
            \put(0.55, 3.00){ \scalebox{1.1}{\colorbox{white}{\makebox{2.31 T}}} }
            \put( 0.0, 3.31){ \scalebox{1.1}{\makebox{new AFM phase}}}
            
            \put(-0.4, 3.20){ \scalebox{1.0}{\color{blue}{\makebox{A}}}}
            \linethickness{0.3 mm}
            \multiput(-0.15,-0.0)(0,0.2){16}{\color{blue}{\line(0,1){0.1}}}
        }
        }
        
        \put(0,0.2){
        \put(0.3, 3.6){
            \put(0.0, 0.0){\includegraphics[width=2.3 cm,trim={26.9cm 24.cm 26.9cm 24.cm},clip=true,frame]{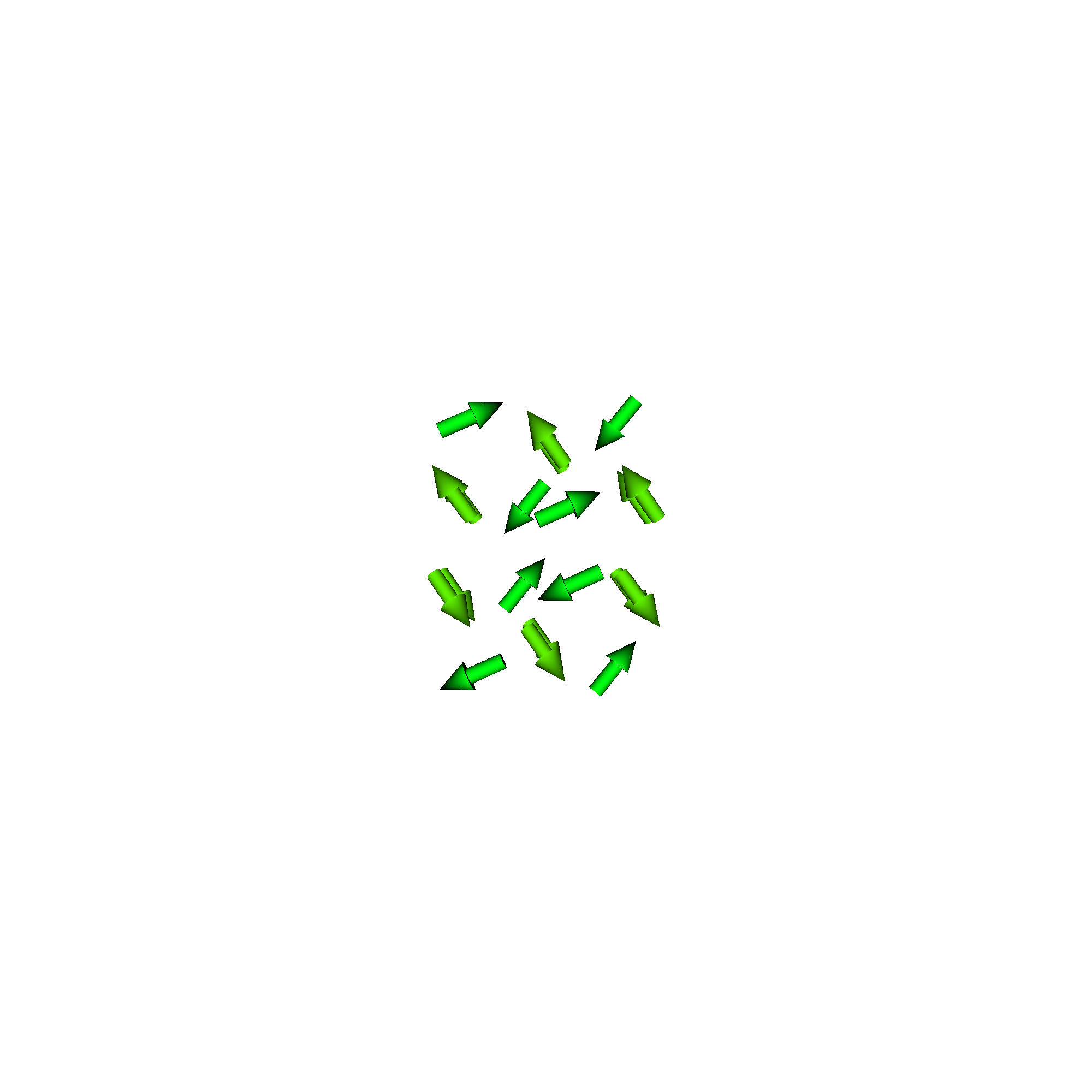}}
            \put(0.55, 3.00){ \scalebox{1.1}{\colorbox{white}{\makebox{2.4 T}}} }
        }
        \put(2.85, 3.6){
            \put(0.0, 0.0){\includegraphics[width=2.3 cm,trim={26.9cm 24.cm 26.9cm 24.cm},clip=true,frame]{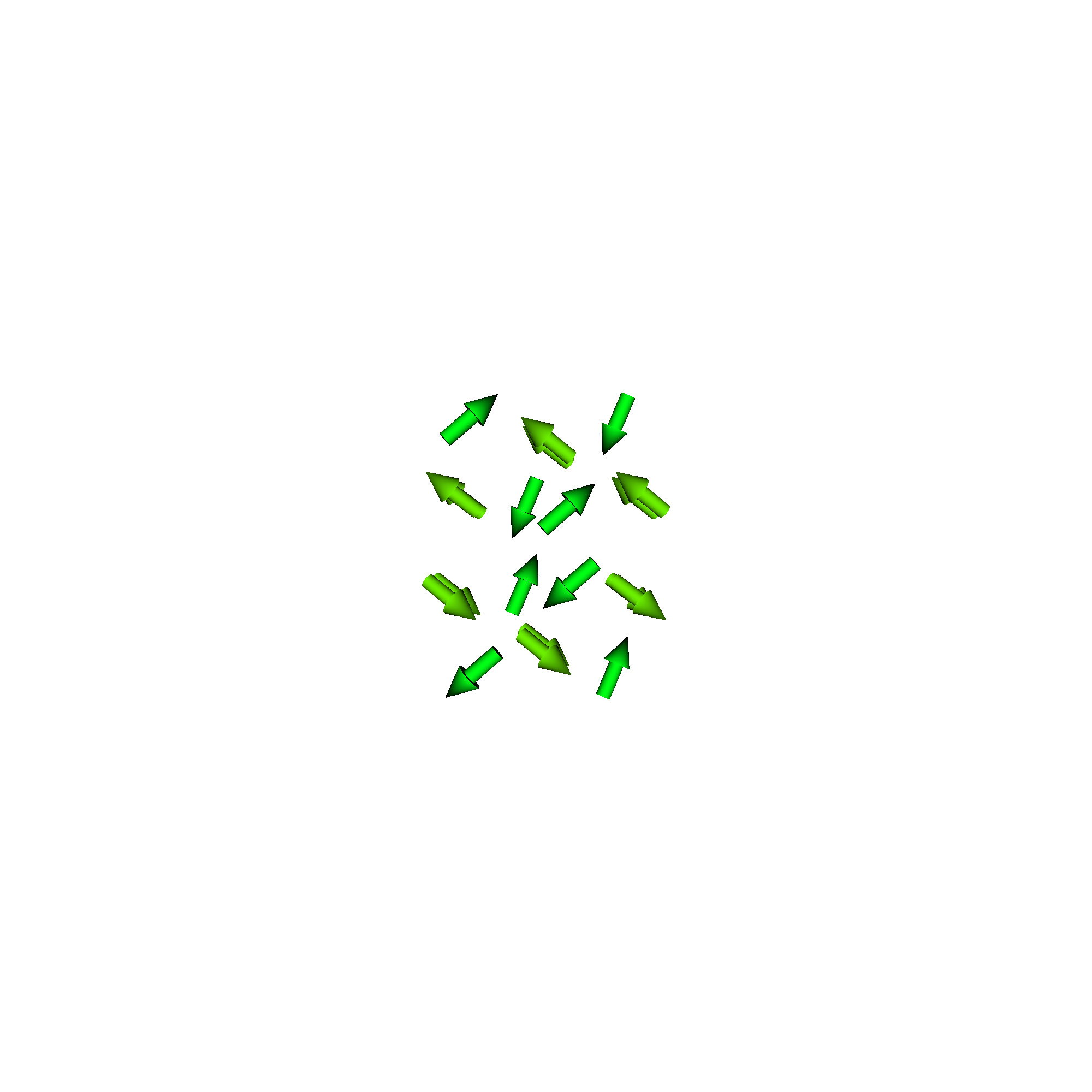}}
            \put(0.55, 3.00){ \scalebox{1.1}{\colorbox{white}{\makebox{2.5 T}}} }
        }
        \put(5.5, 3.6){
            \put(0.0, 0.0){\includegraphics[width=2.3 cm,trim={26.9cm 24.cm 26.9cm 24.cm},clip=true,frame]{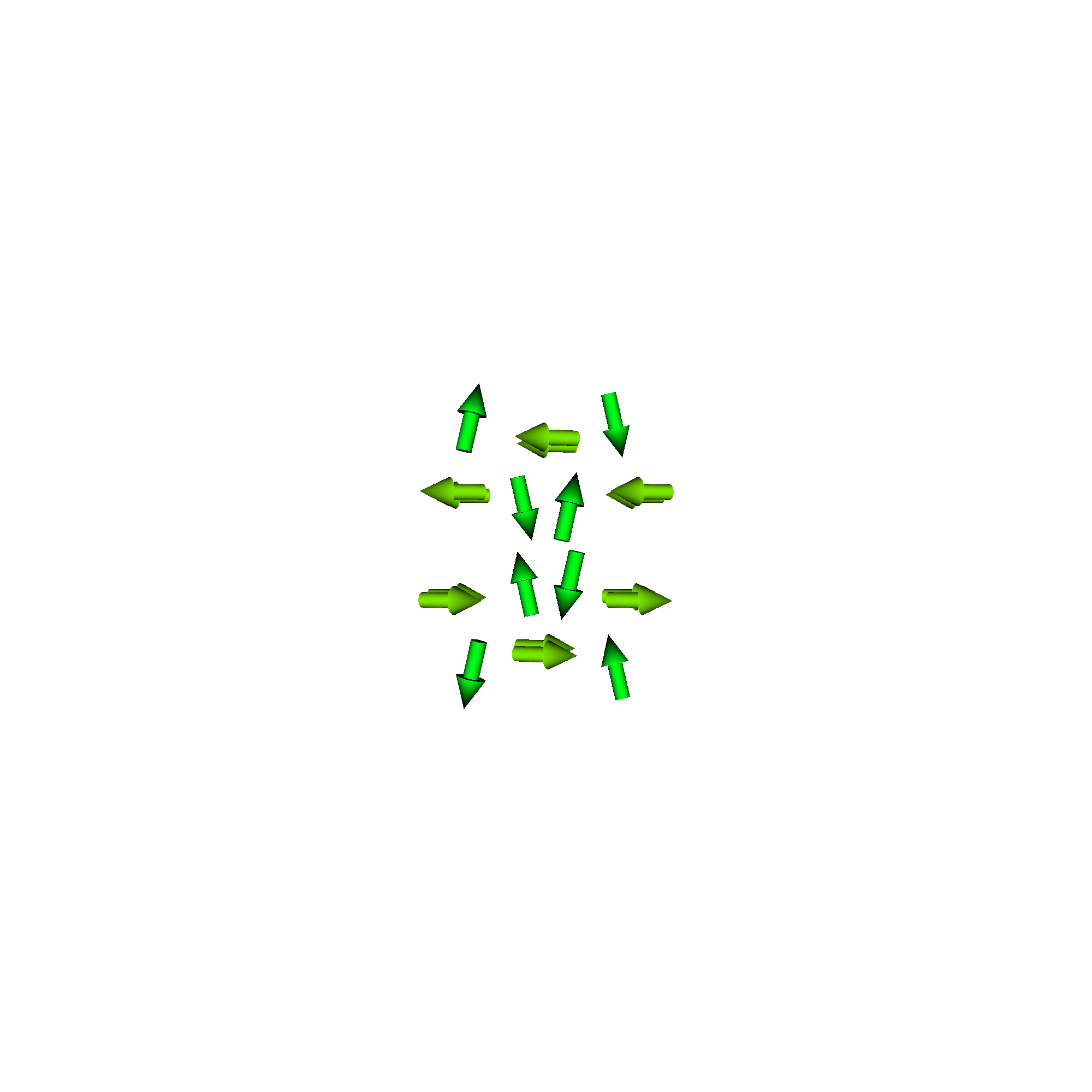}}
            \put(0.55, 3.00){ \scalebox{1.1}{\colorbox{white}{\makebox{2.6 T}}} }
            \put(1.65, 3.20){ \scalebox{1.1}{\makebox{AFM1'}}}
            
            \put(-0.4, 3.20){ \scalebox{1.0}{\color{blue}{\makebox{B}}}}
            \linethickness{0.3 mm}
            \multiput(-0.15,-0.0)(0,0.2){16}{\color{blue}{\line(0,1){0.1}}}
        }
        }
        
        \put(0,0){
        \put(0.3, 0.1){
            \put(0.0, 0.0){\includegraphics[width=2.3 cm,trim={26.9cm 24.cm 26.9cm 24.cm},clip=true,frame]{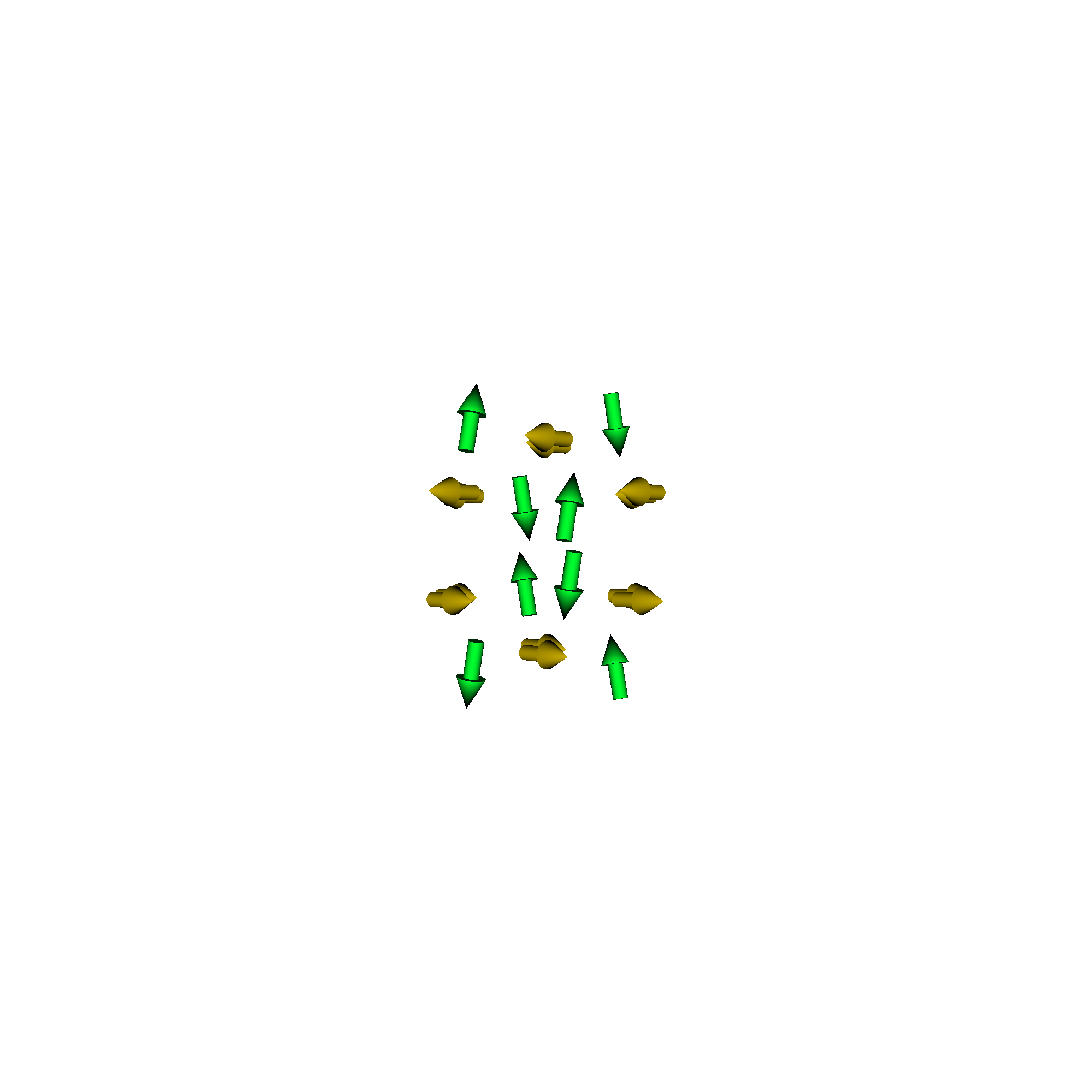}}
            \put(0.55, 3.00){ \scalebox{1.1}{\colorbox{white}{\makebox{5.0 T}}} }
        }
        \put(2.85, 0.1){
            \put(0.0, 0.0){\includegraphics[width=2.3 cm,trim={26.9cm 24.cm 26.9cm 24.cm},clip=true,frame]{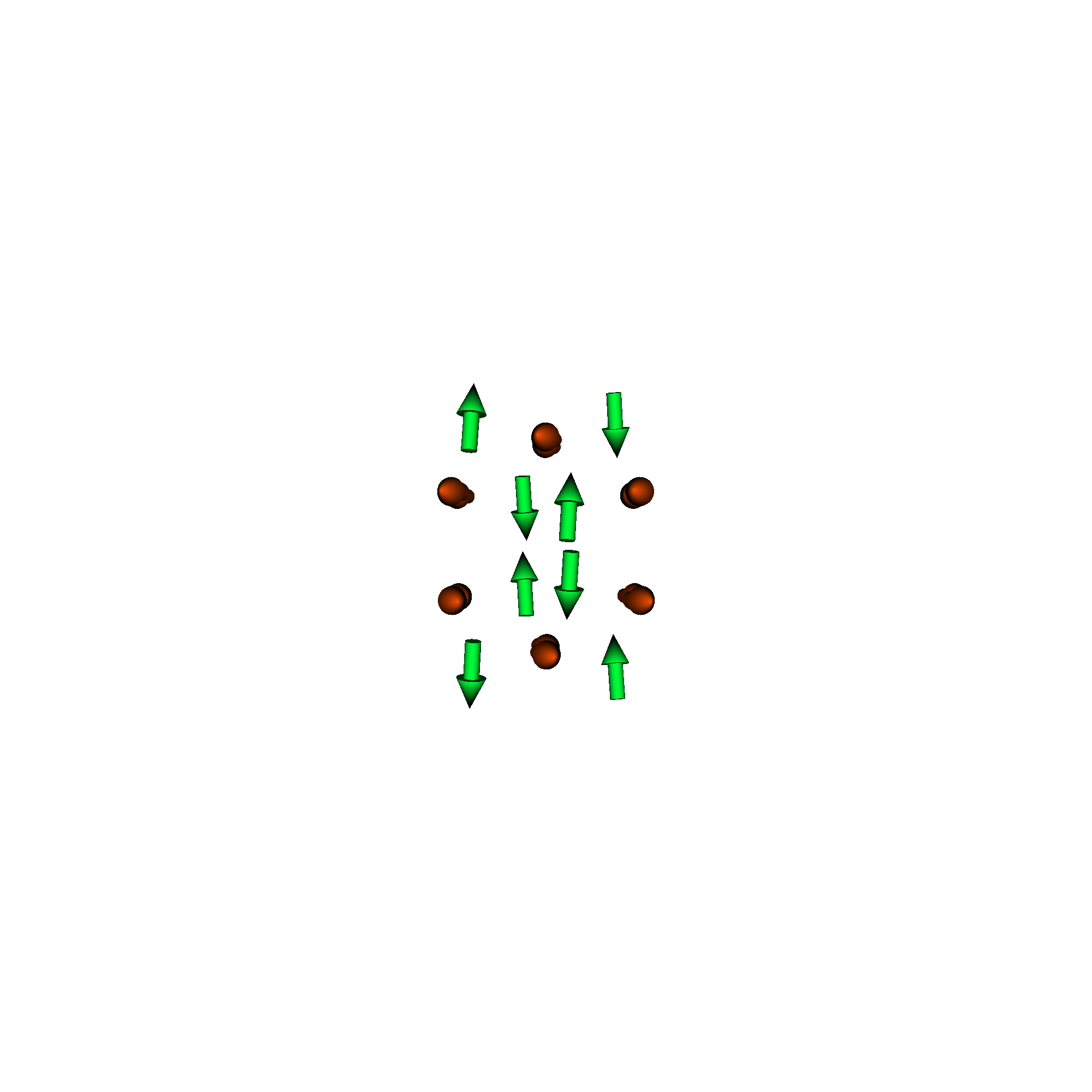}}
            \put(0.55, 3.00){ \scalebox{1.1}{\colorbox{white}{\makebox{6.2 T}}} }
        }
        \put(5.5, 0.1){
            \put(0.0, 0.0){\includegraphics[width=2.3 cm,trim={26.9cm 24.cm 26.9cm 24.cm},clip=true,frame]{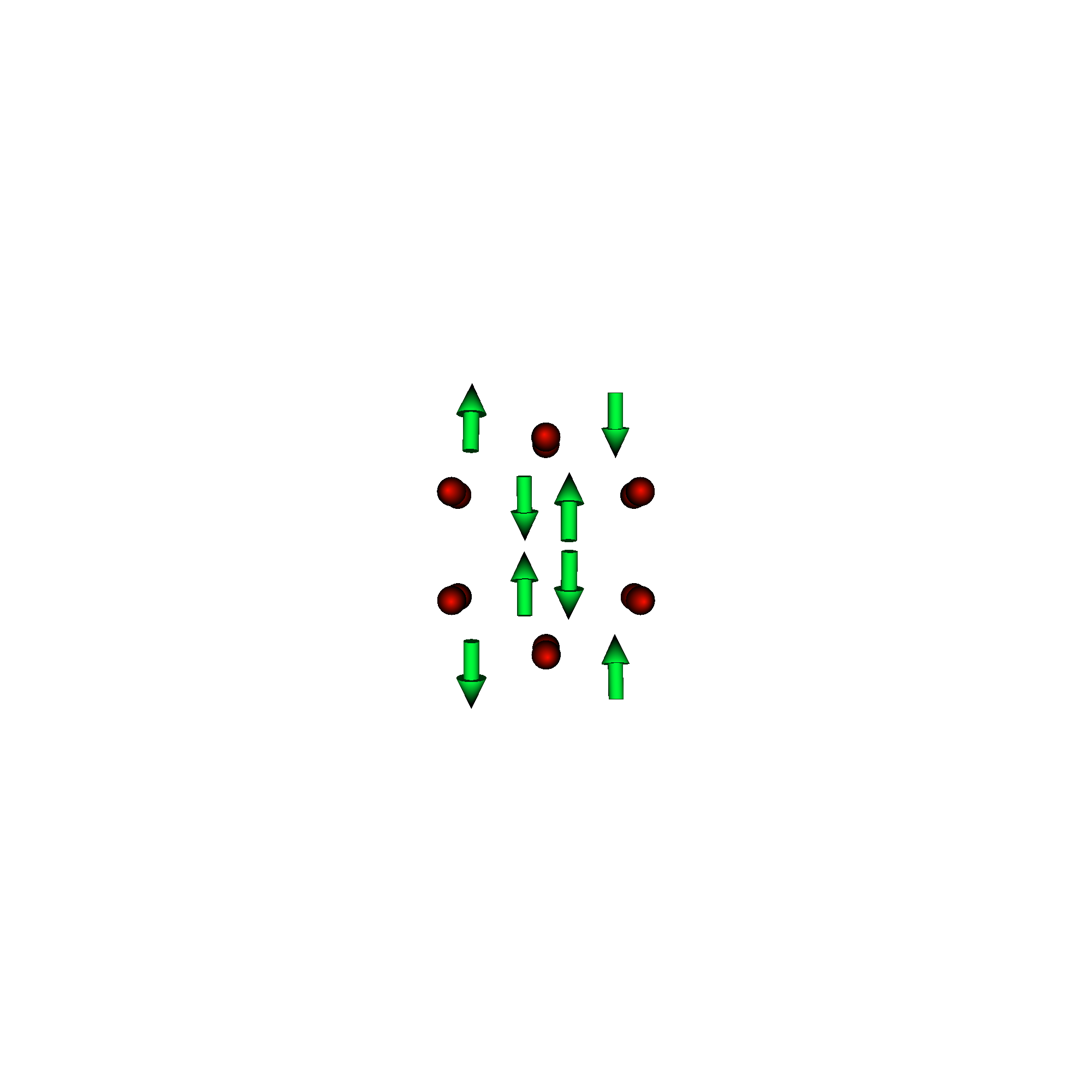}}
            \put(0.55, 3.00){ \scalebox{1.1}{\colorbox{white}{\makebox{6.4 T}}} }
            \put(0.9, 3.35){ \scalebox{1.1}{\makebox{AFM2-like}}}
            
            \put(-0.4, 3.20){ \scalebox{1.0}{\color{blue}{\makebox{C}}}}
            \linethickness{0.3 mm}
            \multiput(-0.15,-0.0)(0,0.2){16}{\color{blue}{\line(0,1){0.1}}}
        }
        }
        
        \put(-0.2,-0.3) {
            \put( 0.05, 0 ){ \includegraphics[width=1.0cm,trim={30cm 29cm 10cm  10cm  },clip=true]{axes.png} }
            \put( 1.1, 0.1){ \scalebox{0.9}{\makebox{a}} }
            \put(   0, 1.1){ \scalebox{0.9}{\makebox{b}} }
            \put(   0, 0.1){ \scalebox{0.9}{\makebox{c}} }
        }
    }
  \end{picture}

  \caption{\label{fig:field_configurations}
  Spin configuration of $\textnormal{Mn}_5\textnormal{Si}_3$ (projected in the $ab$ plane of the orthorhombic cell) at various applied fields for $\VEC B\parallel \hat{\VEC c}$.
  For simplicity we show only the spin orientation (as arrows) of the Mn atoms that have a magnetic moment according to our model.
  The colour scale refers to the in-plane (greenish) or out-of-plane (blue, red, yellow) projection of the spins. Blue and red/yellow arrows point in and out of the page, respectively.
  The dashed lines indicate the possible field-induced phase transitions.
  The figures next to the dashed lines show that spin configuration immediately before and after the phase transition.
  }
\end{figure}

In order to examine further if the theoretically obtained exchange interactions that stabilize a noncollinear spin arrangement for $\textnormal{Mn}_5\textnormal{Si}_3$ (see Fig.~\ref{fig:all_AFM1}(d)) and reproduce the experimentally obtained magnon dispersion (see Figs.~\ref{fig:theory_dispersions}) are realistic, we simulated the evolution of the magnetic structure under an applied magnetic field. To this aim we included to our model Hamiltonian a Zeeman term with the field applied along the $c$ axis, since for this field orientation the magnetic phase diagram is well established (see Fig.~\ref{fig:phase_diagram}).

We started from a random spin configuration and performed spin dynamics simulations to determine an equilibrium state under different magnetic fields.
The same random initial configuration was used for all fields.
After obtaining the equilibrium spin configuration, we calculated the energies of the spin-wave modes.
By analyzing the discontinuities in the evolution of the spin-wave energies as a function of the external field, we identified three phase transitions, marked as A, B, and C in Figs.~\ref{fig:field_configurations}.

In the ground-state ($B = 0$\,T), the spins are noncollinear and lie in the $bc$ plane, which is the most energetically favorable plane when taking into account the magnetocrystalline anisotropies of the system.
In this state, the Mn1 spins are mostly along the easy axis $b$, while the Mn2 spins have significant components along the $c$ axis, see Fig.~\ref{fig:all_AFM1}(d).
Our proposed magnetic ground-state for the AFM1 phase indicates that all the Mn1 and two-thirds of the Mn2 sites carry a magnetic moment. Although this is in agreement with most neutron diffraction data~\cite{gottschilch_study_2012,Lander_1967,brown_low-temperature_1992} a difference is observed regarding the spin orientation of the moments in the AFM1 phase in literature (see Fig.~\ref{fig:all_AFM1}(b)-(d)).

At a magnetic field of $\sim$ 2.31\,T, the system undergoes the first phase transition (labeled A in Fig.~\ref{fig:field_configurations}), which can be imagined as a ``spin-flop" transition for the Mn2 spins where all the spins then lie mostly in the $ab$ plane.
In previous studies, a weak change of the magnetic susceptibility at 30\,K was reported in powder samples~\cite{gottschilch_study_2012} and for $T < 20$\,K a reduction in the magnitude of the anomalous Hall effect was found in single crystals~\cite{surgers_switching_2017}. These observations hinted to a change of the spin configuration of $\textnormal{Mn}_5\textnormal{Si}_3$ for small magnetic fields ($B < 1$\,T) at low temperatures ($T < 30$\,K). 
From powder neutron diffraction studies~\cite{gottschilch_study_2012}, it is still not clear whether this is a new phase or if the features observed so far could be associated with weak rearrangement of the spins on the Mn2 site due to magnetic frustration or magnetic anisotropies.
However, our model indicates a significant rearrangements of spins in all magnetic sites (Mn1 and two-thirds of the Mn2) which results in another noncollinear phase.
This new AFM phase survives in a narrow magnetic field range, which dependents on the second anisotropy parameter $k_c$.
According to our model, if $k_c$ is reduced the critical field for this first phase transition also decreases.

At a magnetic field of about 2.6\,T, a second transition occurs (labeled as B in Fig.~\ref{fig:field_configurations}), which is more subtle.
According to our model, within this phase the spins lie mostly in the $ab$ plane and start to acquire a component along the $c$ axis.
This phase is possibly associated with the reported AFM1' in several studies~\cite{surgers_switching_2017,biniskos_spin_2018,das_observation_2019} which, however, is observed at higher magnetic fields of $\sim$ 5.5\,T at 10\,K in experiments (see Fig.~\ref{fig:phase_diagram}). Similarly with our results, previous investigations indicate that the field-induced phase AFM1' is expected to host a noncollinear magnetic structure, as it has non zero Hall resistivity~\cite{surgers_large_2014,surgers_switching_2017}. A neutron powder diffraction study~\cite{gottschilch_study_2012} under a magnetic field of 4\,T and for temperatures between 5 to 50\,K, suggested a modification of the crystal and magnetic structure, which leads the magnetic order in the Mn1 sites to vanish. However, a sizable field-induced FM component for all magnetic sites is measured along the field direction of about 0.1\,$\mu_\mathrm{B}$/Mn (single crystal neutron diffraction data at $T = 58$\,K and $B = 3$\,T)~\cite{silva_magnetic_2002}, which is in agreement with our simulations regarding the AFM1' phase. 

Finally, a third phase transition takes place (labeled as C in Fig.~\ref{fig:field_configurations}) at $\sim$ 6.4\,T.
This phase relates to the experimentally reported field-induced AFM2 phase~\cite{silva_magnetic_2002,surgers_switching_2017,biniskos_spin_2018,das_observation_2019} and occurs at about 9.5\,T at 10\,K (see Fig.~\ref{fig:phase_diagram}). Magnetization~\cite{Alkanani_1995} and electric transport measurements~\cite{surgers_large_2014,surgers_switching_2017}, as well as inelastic neutron scattering studies~\cite{biniskos_spin_2018}, proposed that the field-induced AFM2 phase exhibits similar properties to the zero field collinear AFM2 phase observed at $60 < T < 100$\,K. Consistently, our model indicates that the Mn2 spins are mostly collinear and antiparallel to each other as in the AFM2 phase illustrated in Fig.~\ref{fig:all_AFM1}(a), with the central difference that the Mn1 sites now host non-vanishing magnetic moments aligned parallel to the magnetic field direction. We note that to our knowledge and until nowadays neutron diffraction studies are performed for magnetic fields less than 5\,T~\cite{gottschilch_study_2012,silva_magnetic_2002} in the temperature range where the transitions take place and therefore, it is not clear if the Mn1 moment collapses in the field-induced AFM2 phase for stronger magnetic fields. 
Concerning the one-third of the Mn2 atoms that have no ordered moment at zero field~\cite{gottschilch_study_2012,Lander_1967,brown_low-temperature_1992} it is suggested that they do not acquire a field-induced moment with field. According to Ref.~\onlinecite{silva_magnetic_2002} the Mn magnetic moments and their magnitude are dependent on the local environment, and an aligned moment in this position (one-third of the Mn2 atoms) may be only attributed to the presence of a local field produced by aligned moments on neighbouring magnetic atoms (two-thirds of the Mn2 atoms), instead of a direct effect of the magnetic field on a local magnetic moment.

Our results show that the spin texture of $\textnormal{Mn}_5\textnormal{Si}_3$ under external magnetic field consists of a non trivial AFM alignment of the Mn spins. 
As in the cases of $\textnormal{Mn}_3\textnormal{Sn}$~\cite{Nakatsuji_2015,Ikhlas_2017} and $\textnormal{Mn}_3\textnormal{Ge}$~\cite{Kiyohara_2016,Wuttke_2019}, one would expect that the noncollinearity of the Mn moments in $\textnormal{Mn}_5\textnormal{Si}_3$ apart from the already discovered large anomalous Hall conductivity~\cite{surgers_large_2014} could also generate an anomalous Nernst effect.  
The observation of an anomalous Nernst effect with thermotransport measurements, so far not reported for $\textnormal{Mn}_5\textnormal{Si}_3$, would be extremely useful to provide a measure of the Berry curvature at the Fermi level~\cite{Xiao_2006} and to pave the way for further studies in search of topological signatures.

\subsection{Susceptible Mn1 moments under a magnetic field}

We have shown that our proposed model for the noncollinear AFM1 phase of $\textnormal{Mn}_5\textnormal{Si}_3$ manages to qualitatively reproduce all the phase transitions observed experimentally in the $B-T$ phase diagram. 
A central assumption of the Heisenberg model is that the magnetic moments are rigid against longitudinal variations, that is, their size is unchangeable.
This is reasonable for the Mn2 sites while being more questionable for the Mn1 sites.
In spite of this, we showed that one can stabilise a collinear AFM2-like phase by applying an external magnetic field while considered rigid Mn1 magnetic moments.
One open question is whether the size of these moments can be affected by the influence of an external field and what is the mechanism behind that.

We assume that the Mn1 magnetic moments are longitudinally susceptible, i.e., the energy cost to change their length is comparable to the other energy scales, such as the exchange and Zeeman energies.
Then, it could be energetically favorable for the system to collapse a magnetic moment instead of having to deal with a group of three or more moments frustrated due to the exchange interaction.

\begin{figure}[tb]
  \setlength{\unitlength}{1cm}
  \newcommand{\boxsize}{0.3}
  \begin{picture}(5,3)
      \put( 0.0, 0.0){ \includegraphics[width=3.7cm, trim={0.0cm 1.4cm 0cm 0cm},clip=true]{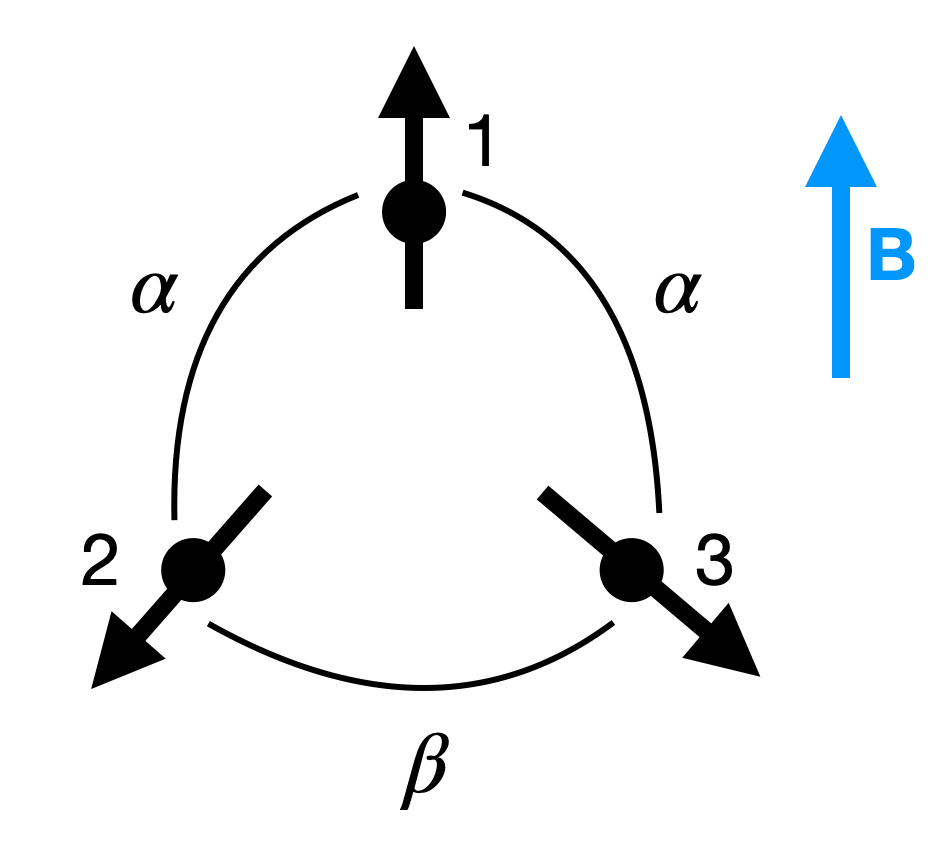} } 
  \end{picture}
  \caption{\label{fig:scheme}
Model three-spin system.
The spin in site 1 is susceptible to variations in its spin length.
An external field is applied parallel to $S_1$.
  }
\end{figure}

To test this hypothesis, we consider a system of three spins in an equilateral triangle coupled antiferromagnetically, see Fig.~\ref{fig:scheme}.
First, we recapitulate the results for the case in which all three spins and their mutual interactions are equivalent.
The exchange interaction is described by:
\begin{equation}
    \mathcal{H}_J = - \frac{1}{2} \sum_{ij} J_{ij} \VEC{S}_i \cdot \VEC{S}_j
    = JS^2(2\cos{\alpha} + \cos{\beta} ) ,
\end{equation}
where $\alpha$ and $\beta$ the angles between the corresponding spins and we considered that $J_{ij}=-J$ ( with $J>0$) for all interactions.
Thus, the exchange energy of a FM alignment is $3 J S^2$; for a state with two parallel spins and another antiparallel, the energy is $-J S^2$; and the energy of a $120^\circ$-state (Fig.~\ref{fig:scheme} with $\alpha=\beta=120^\circ$) is $-3 J S^2/2$. The latter state is clearly more favorable than the other two states and is therefore the ground-state.

Now we relate this model to the situation found in $\textnormal{Mn}_5\textnormal{Si}_3$ (see the black triangle in Fig.~\ref{fig:all_AFM1}(d)).
We consider that two magnetic moments are rigid with respect to longitudinal variations (in analogy to the Mn2 moments in $\textnormal{Mn}_5\textnormal{Si}_3$), while the third moment (for example on site 1 in Fig.~\ref{fig:scheme} relates to the Mn1 moment in $\textnormal{Mn}_5\textnormal{Si}_3$) is susceptible and we model its potential energy with the simplest quadratic form:

\begin{equation}
\mathcal E(S_1) = \frac{1}{2\chi} (S_1 - S_0)^2 ,
\end{equation}
with $\chi>0$, such that there is an energy cost to increase or decrease $S_1$ from a reference value $S_0$.
The rigid spin limit is obtained for $\chi \rightarrow 0$.
Furthermore, we consider that the coupling between $S_1$ and the other two spins is given by $J'>0$ and can be different from $J$ that couples $S_2$ to $S_3$.
We apply an in-plane magnetic field $\VEC B$ parallel to $\VEC S_1$, as in Fig.~\ref{fig:scheme} and then the Hamiltonian reads:
\begin{align}
    \mathcal H &= 2J' S S_1 \cos{\alpha} + J S^2 \cos{2 \alpha} \nonumber\\
    &+ \frac{1}{2\chi} \left(S_1-S_0\right)^2 - B S_1 - 2 B S \cos \alpha \;,
\end{align}
where $S_2=S_3=S$, and we used the relation $2\alpha + \beta = 2 \pi$ (the spins are assumed to be always in the same plane).

By minimizing the Hamiltonian with respect to $S_1$, one obtains that:
\begin{equation}
    \frac{\partial\mathcal H}{\partial S_1} = 0 \;\;\rightarrow\;\; S_1 = S_0 + \chi \left(B - 2J' S \cos{\alpha}\right) .
\end{equation}
With respect to $\alpha$, the stationary condition gives:
\begin{equation}
    \frac{\partial\mathcal H}{\partial \alpha} = 0 \;\;\rightarrow\;\; \cos\alpha = \frac{B - J' S_1}{2JS} \quad\text{or}\quad \sin\alpha = 0 \;.
\end{equation}
The first solution minimizes the Hamiltonian while the second, $\sin\alpha = 0$, gives two collinear states with spins 2 and 3 parallel to each other maximizing the Hamiltonian.
Assuming that $\sin\alpha \neq 0$, we find:
\begin{align}
    S_1 &= a \left(S_0 + b \chi  B\right) \nonumber \quad \textnormal{where,} \\
    a =& \left(1 - \chi \frac{(J')^2}{J}\right)^{-1} \quad \textnormal{and} \quad
    b = \left(1 - \frac{J'}{J}\right) .
\end{align}
We are assuming that $J, J', \chi > 0$, so $a$ diverges for $\chi = J/(J')^2$; and $b=0$ for $J'=J$.
Let first consider the case when $a$ is positive ($\chi < J/(J')^2$).
In this case, $S_1$ is finite at zero field.
For $J' > J$ ($b<0$), $S_1$ reduces linearly with the field; it remains constant if $J'=J$; and it increases for $J'<J$.
Another interesting scenario is for $\chi \rightarrow \infty$ ($a\rightarrow0$).
In this case, $S_1$ vanishes for zero field.
Then, it increases parallel or antiparallel to the field depending on the sign of $b$.

\section{Conclusions}

By combining inelastic neutron scattering measurements and density functional theory calculations we investigated the magnetic ground-state and spin dynamics of the AFM1 phase of $\textnormal{Mn}_5\textnormal{Si}_3$. 
The experimentally obtained spin excitation spectrum along the three main crystal axes of the orthorhombic symmetry at $T = 10$\,K is characterized by a spin gap of the order of 1\,meV, steep magnon dispersions and a low optic magnon mode that originates at about 10\,meV. The INS results can be well described by a Heisenberg Hamiltonian using six exchange interactions.
In addition, our theoretical model suggests a noncollinear magnetic ground-state for $\textnormal{Mn}_5\textnormal{Si}_3$ which is different from all the previous proposed ones based on neutron diffraction data. 
According to our model the Mn1 spins are aligned mainly along the easy axis $b$, while the Mn2 spins have components in the $bc$ plane, which is a result of the magnetocrystalline anisotropies of the system, that makes the $bc$ plane the most energetically favorable one.
The existing controversy in literature and our new results demonstrates that the magnetic ground state of $\textnormal{Mn}_5\textnormal{Si}_3$ needs to be reexamined by employing modern polarized neutron diffraction techniques. 

Applying an external magnetic field parallel to the $c$ axis of the orthorhombic cell results in  field-induced transitions that are overall in qualitative agreement with the experimentally established $B-T$ phase diagram.
Furthermore, our work supports the scenario of another AFM phase at weak magnetic fields that has been hinted to exist in previous studies. However, a clear phase boundary between this new phase and the already confirmed AFM phases needs to be evidenced experimentally by magnetization measurements on single crystals. Also neutron diffraction studies on single crystals for different magnetic fields at base temperature would be highly desirable in order to investigate experimentally the spin texture of all field-induced transitions and compare them with our results.

Finally, we attempt to elucidate the mechanism behind the establishment of a collinear AFM2-like spin arrangement due to an external magnetic field from a highly noncolinear AFM1 phase at zero field. 
To this aim, we employ a theoretical model where we examine the stability of a magnetic moment in a frustrated spin arrangement and under the influence of a magnetic field. 
The proposed model demonstrates that if one moment is sensitive to longitudinal variations then it can change its magnitude under an external magnetic field.
Therefore, our model supports the scenario where $\textnormal{Mn}_5\textnormal{Si}_3$ under magnetic field can acquire a collinear AFM2-like spin arrangement for the Mn2 spins that coexist with non vanishing Mn1 moments.

\section{Acknowledgments}

N.B. and F.J.d.S. contributed equally to this work.
N.B. acknowledges the support of JCNS through the Tasso Springer fellowship. 
F.J.d.S. acknowledges support of the European H2020 Intersect project (Grant No. 814487), and N.M. of the Swiss National Science Foundation (SNSF) through its National Centre of Competence in Research (NCCR) MARVEL.
This work was also supported by the Brazilian funding agency CAPES under Project No. 13703/13-7 and the European Research Council (ERC) under the European Union's Horizon 2020 research and innovation program (ERC-consolidator Grant No. 681405-DYNASORE). We gratefully acknowledge the computing time granted by JARA-HPC on the supercomputer JURECA at Forschungszentrum J\"ulich and by RWTH Aachen University.

\bibliography{paper_Mn5Si3}

\begin{thebibliography}{42}%
\makeatletter
\providecommand \@ifxundefined [1]{%
 \@ifx{#1\undefined}
}%
\providecommand \@ifnum [1]{%
 \ifnum #1\expandafter \@firstoftwo
 \else \expandafter \@secondoftwo
 \fi
}%
\providecommand \@ifx [1]{%
 \ifx #1\expandafter \@firstoftwo
 \else \expandafter \@secondoftwo
 \fi
}%
\providecommand \natexlab [1]{#1}%
\providecommand \enquote  [1]{``#1''}%
\providecommand \bibnamefont  [1]{#1}%
\providecommand \bibfnamefont [1]{#1}%
\providecommand \citenamefont [1]{#1}%
\providecommand \href@noop [0]{\@secondoftwo}%
\providecommand \href [0]{\begingroup \@sanitize@url \@href}%
\providecommand \@href[1]{\@@startlink{#1}\@@href}%
\providecommand \@@href[1]{\endgroup#1\@@endlink}%
\providecommand \@sanitize@url [0]{\catcode `\\12\catcode `\$12\catcode
  `\&12\catcode `\#12\catcode `\^12\catcode `\_12\catcode `\%12\relax}%
\providecommand \@@startlink[1]{}%
\providecommand \@@endlink[0]{}%
\providecommand \url  [0]{\begingroup\@sanitize@url \@url }%
\providecommand \@url [1]{\endgroup\@href {#1}{\urlprefix }}%
\providecommand \urlprefix  [0]{URL }%
\providecommand \Eprint [0]{\href }%
\providecommand \doibase [0]{http://dx.doi.org/}%
\providecommand \selectlanguage [0]{\@gobble}%
\providecommand \bibinfo  [0]{\@secondoftwo}%
\providecommand \bibfield  [0]{\@secondoftwo}%
\providecommand \translation [1]{[#1]}%
\providecommand \BibitemOpen [0]{}%
\providecommand \bibitemStop [0]{}%
\providecommand \bibitemNoStop [0]{.\EOS\space}%
\providecommand \EOS [0]{\spacefactor3000\relax}%
\providecommand \BibitemShut  [1]{\csname bibitem#1\endcsname}%
\let\auto@bib@innerbib\@empty
\bibitem [{\citenamefont {Ikhlas}\ \emph {et~al.}(2017)\citenamefont {Ikhlas},
  \citenamefont {Tomita}, \citenamefont {Koretsune}, \citenamefont {Suzuki},
  \citenamefont {Nishio-Hamane}, \citenamefont {Arita}, \citenamefont {Otani},\
  and\ \citenamefont {Nakatsuji}}]{Ikhlas_2017}%
  \BibitemOpen
  \bibfield  {author} {\bibinfo {author} {\bibfnamefont {M.}~\bibnamefont
  {Ikhlas}}, \bibinfo {author} {\bibfnamefont {T.}~\bibnamefont {Tomita}},
  \bibinfo {author} {\bibfnamefont {T.}~\bibnamefont {Koretsune}}, \bibinfo
  {author} {\bibfnamefont {M.-T.}\ \bibnamefont {Suzuki}}, \bibinfo {author}
  {\bibfnamefont {D.}~\bibnamefont {Nishio-Hamane}}, \bibinfo {author}
  {\bibfnamefont {R.}~\bibnamefont {Arita}}, \bibinfo {author} {\bibfnamefont
  {Y.}~\bibnamefont {Otani}}, \ and\ \bibinfo {author} {\bibfnamefont
  {S.}~\bibnamefont {Nakatsuji}},\ }\href {\doibase 10.1038/nphys4181}
  {\bibfield  {journal} {\bibinfo  {journal} {Nat. Phys.}\ }\textbf {\bibinfo
  {volume} {13}},\ \bibinfo {pages} {1085} (\bibinfo {year}
  {2017})}\BibitemShut {NoStop}%
\bibitem [{\citenamefont {Nakatsuji}\ \emph {et~al.}(2015)\citenamefont
  {Nakatsuji}, \citenamefont {Kiyohara},\ and\ \citenamefont
  {Higo}}]{Nakatsuji_2015}%
  \BibitemOpen
  \bibfield  {author} {\bibinfo {author} {\bibfnamefont {S.}~\bibnamefont
  {Nakatsuji}}, \bibinfo {author} {\bibfnamefont {N.}~\bibnamefont {Kiyohara}},
  \ and\ \bibinfo {author} {\bibfnamefont {T.}~\bibnamefont {Higo}},\ }\href
  {\doibase 10.1038/nature15723} {\bibfield  {journal} {\bibinfo  {journal}
  {Nature}\ }\textbf {\bibinfo {volume} {527}},\ \bibinfo {pages} {212}
  (\bibinfo {year} {2015})}\BibitemShut {NoStop}%
\bibitem [{\citenamefont {Kiyohara}\ \emph {et~al.}(2016)\citenamefont
  {Kiyohara}, \citenamefont {Tomita},\ and\ \citenamefont
  {Nakatsuji}}]{Kiyohara_2016}%
  \BibitemOpen
  \bibfield  {author} {\bibinfo {author} {\bibfnamefont {N.}~\bibnamefont
  {Kiyohara}}, \bibinfo {author} {\bibfnamefont {T.}~\bibnamefont {Tomita}}, \
  and\ \bibinfo {author} {\bibfnamefont {S.}~\bibnamefont {Nakatsuji}},\ }\href
  {\doibase 10.1103/PhysRevApplied.5.064009} {\bibfield  {journal} {\bibinfo
  {journal} {Phys. Rev. Applied}\ }\textbf {\bibinfo {volume} {5}},\ \bibinfo
  {pages} {064009} (\bibinfo {year} {2016})}\BibitemShut {NoStop}%
\bibitem [{\citenamefont {Park}\ \emph {et~al.}(2018)\citenamefont {Park},
  \citenamefont {Oh}, \citenamefont {Uhlířová}, \citenamefont {Jackson},
  \citenamefont {Deák}, \citenamefont {Szunyogh}, \citenamefont {Lee},
  \citenamefont {Cho}, \citenamefont {Kim}, \citenamefont {Walker},
  \citenamefont {Adroja}, \citenamefont {Sechovský},\ and\ \citenamefont
  {Park}}]{Park_2018}%
  \BibitemOpen
  \bibfield  {author} {\bibinfo {author} {\bibfnamefont {P.}~\bibnamefont
  {Park}}, \bibinfo {author} {\bibfnamefont {J.}~\bibnamefont {Oh}}, \bibinfo
  {author} {\bibfnamefont {K.}~\bibnamefont {Uhlířová}}, \bibinfo {author}
  {\bibfnamefont {J.}~\bibnamefont {Jackson}}, \bibinfo {author} {\bibfnamefont
  {A.}~\bibnamefont {Deák}}, \bibinfo {author} {\bibfnamefont
  {L.}~\bibnamefont {Szunyogh}}, \bibinfo {author} {\bibfnamefont {K.~H.}\
  \bibnamefont {Lee}}, \bibinfo {author} {\bibfnamefont {H.}~\bibnamefont
  {Cho}}, \bibinfo {author} {\bibfnamefont {H.-L.}\ \bibnamefont {Kim}},
  \bibinfo {author} {\bibfnamefont {H.~C.}\ \bibnamefont {Walker}}, \bibinfo
  {author} {\bibfnamefont {D.}~\bibnamefont {Adroja}}, \bibinfo {author}
  {\bibfnamefont {V.}~\bibnamefont {Sechovský}}, \ and\ \bibinfo {author}
  {\bibfnamefont {J.-G.}\ \bibnamefont {Park}},\ }\href {\doibase
  10.1038/s41535-018-0137-9} {\bibfield  {journal} {\bibinfo  {journal} {npj
  Quantum Materials}\ }\textbf {\bibinfo {volume} {3}},\ \bibinfo {pages} {63}
  (\bibinfo {year} {2018})}\BibitemShut {NoStop}%
\bibitem [{\citenamefont {Sukhanov}\ \emph {et~al.}(2019)\citenamefont
  {Sukhanov}, \citenamefont {Pavlovskii}, \citenamefont {Bourges},
  \citenamefont {Walker}, \citenamefont {Manna}, \citenamefont {Felser},\ and\
  \citenamefont {Inosov}}]{Sukhanov_2019}%
  \BibitemOpen
  \bibfield  {author} {\bibinfo {author} {\bibfnamefont {A.~S.}\ \bibnamefont
  {Sukhanov}}, \bibinfo {author} {\bibfnamefont {M.~S.}\ \bibnamefont
  {Pavlovskii}}, \bibinfo {author} {\bibfnamefont {P.}~\bibnamefont {Bourges}},
  \bibinfo {author} {\bibfnamefont {H.~C.}\ \bibnamefont {Walker}}, \bibinfo
  {author} {\bibfnamefont {K.}~\bibnamefont {Manna}}, \bibinfo {author}
  {\bibfnamefont {C.}~\bibnamefont {Felser}}, \ and\ \bibinfo {author}
  {\bibfnamefont {D.~S.}\ \bibnamefont {Inosov}},\ }\href {\doibase
  10.1103/PhysRevB.99.214445} {\bibfield  {journal} {\bibinfo  {journal} {Phys.
  Rev. B}\ }\textbf {\bibinfo {volume} {99}},\ \bibinfo {pages} {214445}
  (\bibinfo {year} {2019})}\BibitemShut {NoStop}%
\bibitem [{\citenamefont {Chen}\ \emph {et~al.}(2020)\citenamefont {Chen},
  \citenamefont {Gaudet}, \citenamefont {Dasgupta}, \citenamefont {Marcus},
  \citenamefont {Lin}, \citenamefont {Chen}, \citenamefont {Tomita},
  \citenamefont {Ikhlas}, \citenamefont {Zhao}, \citenamefont {Chen},
  \citenamefont {Stone}, \citenamefont {Tchernyshyov}, \citenamefont
  {Nakatsuji},\ and\ \citenamefont {Broholm}}]{Chen_2020}%
  \BibitemOpen
  \bibfield  {author} {\bibinfo {author} {\bibfnamefont {Y.}~\bibnamefont
  {Chen}}, \bibinfo {author} {\bibfnamefont {J.}~\bibnamefont {Gaudet}},
  \bibinfo {author} {\bibfnamefont {S.}~\bibnamefont {Dasgupta}}, \bibinfo
  {author} {\bibfnamefont {G.~G.}\ \bibnamefont {Marcus}}, \bibinfo {author}
  {\bibfnamefont {J.}~\bibnamefont {Lin}}, \bibinfo {author} {\bibfnamefont
  {T.}~\bibnamefont {Chen}}, \bibinfo {author} {\bibfnamefont {T.}~\bibnamefont
  {Tomita}}, \bibinfo {author} {\bibfnamefont {M.}~\bibnamefont {Ikhlas}},
  \bibinfo {author} {\bibfnamefont {Y.}~\bibnamefont {Zhao}}, \bibinfo {author}
  {\bibfnamefont {W.~C.}\ \bibnamefont {Chen}}, \bibinfo {author}
  {\bibfnamefont {M.~B.}\ \bibnamefont {Stone}}, \bibinfo {author}
  {\bibfnamefont {O.}~\bibnamefont {Tchernyshyov}}, \bibinfo {author}
  {\bibfnamefont {S.}~\bibnamefont {Nakatsuji}}, \ and\ \bibinfo {author}
  {\bibfnamefont {C.}~\bibnamefont {Broholm}},\ }\href {\doibase
  10.1103/PhysRevB.102.054403} {\bibfield  {journal} {\bibinfo  {journal}
  {Phys. Rev. B}\ }\textbf {\bibinfo {volume} {102}},\ \bibinfo {pages}
  {054403} (\bibinfo {year} {2020})}\BibitemShut {NoStop}%
\bibitem [{\citenamefont {Songlin}\ \emph {et~al.}(2002)\citenamefont
  {Songlin}, \citenamefont {Tegus}, \citenamefont {Brück}, \citenamefont
  {Klaasse}, \citenamefont {de~Boer},\ and\ \citenamefont
  {Buschow}}]{songlin_magnetic_2002}%
  \BibitemOpen
  \bibfield  {author} {\bibinfo {author} {\bibfnamefont {D.}~\bibnamefont
  {Songlin}}, \bibinfo {author} {\bibfnamefont {O.}~\bibnamefont {Tegus}},
  \bibinfo {author} {\bibfnamefont {E.}~\bibnamefont {Brück}}, \bibinfo
  {author} {\bibfnamefont {J.~C.~P.}\ \bibnamefont {Klaasse}}, \bibinfo
  {author} {\bibfnamefont {F.~R.}\ \bibnamefont {de~Boer}}, \ and\ \bibinfo
  {author} {\bibfnamefont {K.~H.~J.}\ \bibnamefont {Buschow}},\ }\href
  {\doibase 10.1016/S0925-8388(01)01776-5} {\bibfield  {journal} {\bibinfo
  {journal} {Journal of Alloys and Compounds}\ }\textbf {\bibinfo {volume}
  {334}},\ \bibinfo {pages} {249} (\bibinfo {year} {2002})}\BibitemShut
  {NoStop}%
\bibitem [{\citenamefont {Biniskos}\ \emph {et~al.}(2018)\citenamefont
  {Biniskos}, \citenamefont {Schmalzl}, \citenamefont {Raymond}, \citenamefont
  {Petit}, \citenamefont {Steffens}, \citenamefont {Persson},\ and\
  \citenamefont {Br\"uckel}}]{biniskos_spin_2018}%
  \BibitemOpen
  \bibfield  {author} {\bibinfo {author} {\bibfnamefont {N.}~\bibnamefont
  {Biniskos}}, \bibinfo {author} {\bibfnamefont {K.}~\bibnamefont {Schmalzl}},
  \bibinfo {author} {\bibfnamefont {S.}~\bibnamefont {Raymond}}, \bibinfo
  {author} {\bibfnamefont {S.}~\bibnamefont {Petit}}, \bibinfo {author}
  {\bibfnamefont {P.}~\bibnamefont {Steffens}}, \bibinfo {author}
  {\bibfnamefont {J.}~\bibnamefont {Persson}}, \ and\ \bibinfo {author}
  {\bibfnamefont {T.}~\bibnamefont {Br\"uckel}},\ }\href {\doibase
  10.1103/PhysRevLett.120.257205} {\bibfield  {journal} {\bibinfo  {journal}
  {Physical Review Letters}\ }\textbf {\bibinfo {volume} {120}},\ \bibinfo
  {pages} {257205} (\bibinfo {year} {2018})}\BibitemShut {NoStop}%
\bibitem [{\citenamefont {Das}\ \emph {et~al.}(2019)\citenamefont {Das},
  \citenamefont {Mandal}, \citenamefont {Dutta}, \citenamefont {Pramanick},\
  and\ \citenamefont {Chatterjee}}]{das_observation_2019}%
  \BibitemOpen
  \bibfield  {author} {\bibinfo {author} {\bibfnamefont {S.~C.}\ \bibnamefont
  {Das}}, \bibinfo {author} {\bibfnamefont {K.}~\bibnamefont {Mandal}},
  \bibinfo {author} {\bibfnamefont {P.}~\bibnamefont {Dutta}}, \bibinfo
  {author} {\bibfnamefont {S.}~\bibnamefont {Pramanick}}, \ and\ \bibinfo
  {author} {\bibfnamefont {S.}~\bibnamefont {Chatterjee}},\ }\href {\doibase
  10.1103/PhysRevB.100.024409} {\bibfield  {journal} {\bibinfo  {journal}
  {Physical Review B}\ }\textbf {\bibinfo {volume} {100}},\ \bibinfo {pages}
  {024409} (\bibinfo {year} {2019})},\ \bibinfo {note} {publisher: American
  Physical Society}\BibitemShut {NoStop}%
\bibitem [{\citenamefont {Sürgers}\ \emph {et~al.}(2014)\citenamefont
  {Sürgers}, \citenamefont {Fischer}, \citenamefont {Winkel},\ and\
  \citenamefont {Löhneysen}}]{surgers_large_2014}%
  \BibitemOpen
  \bibfield  {author} {\bibinfo {author} {\bibfnamefont {C.}~\bibnamefont
  {Sürgers}}, \bibinfo {author} {\bibfnamefont {G.}~\bibnamefont {Fischer}},
  \bibinfo {author} {\bibfnamefont {P.}~\bibnamefont {Winkel}}, \ and\ \bibinfo
  {author} {\bibfnamefont {H.~v.}\ \bibnamefont {Löhneysen}},\ }\href
  {\doibase 10.1038/ncomms4400} {\bibfield  {journal} {\bibinfo  {journal}
  {Nature Communications}\ }\textbf {\bibinfo {volume} {5}},\ \bibinfo {pages}
  {3400} (\bibinfo {year} {2014})}\BibitemShut {NoStop}%
\bibitem [{\citenamefont {Sürgers}\ \emph {et~al.}(2016)\citenamefont
  {Sürgers}, \citenamefont {Kittler}, \citenamefont {Wolf},\ and\
  \citenamefont {Löhneysen}}]{Suergers_2016}%
  \BibitemOpen
  \bibfield  {author} {\bibinfo {author} {\bibfnamefont {C.}~\bibnamefont
  {Sürgers}}, \bibinfo {author} {\bibfnamefont {W.}~\bibnamefont {Kittler}},
  \bibinfo {author} {\bibfnamefont {T.}~\bibnamefont {Wolf}}, \ and\ \bibinfo
  {author} {\bibfnamefont {H.~v.}\ \bibnamefont {Löhneysen}},\ }\href
  {\doibase 10.1063/1.4943759} {\bibfield  {journal} {\bibinfo  {journal} {AIP
  Advances}\ }\textbf {\bibinfo {volume} {6}},\ \bibinfo {pages} {055604}
  (\bibinfo {year} {2016})},\ \Eprint
  {http://arxiv.org/abs/https://doi.org/10.1063/1.4943759}
  {https://doi.org/10.1063/1.4943759} \BibitemShut {NoStop}%
\bibitem [{\citenamefont {Sürgers}\ \emph {et~al.}(2017)\citenamefont
  {Sürgers}, \citenamefont {Wolf}, \citenamefont {Adelmann}, \citenamefont
  {Kittler}, \citenamefont {Fischer},\ and\ \citenamefont
  {Löhneysen}}]{surgers_switching_2017}%
  \BibitemOpen
  \bibfield  {author} {\bibinfo {author} {\bibfnamefont {C.}~\bibnamefont
  {Sürgers}}, \bibinfo {author} {\bibfnamefont {T.}~\bibnamefont {Wolf}},
  \bibinfo {author} {\bibfnamefont {P.}~\bibnamefont {Adelmann}}, \bibinfo
  {author} {\bibfnamefont {W.}~\bibnamefont {Kittler}}, \bibinfo {author}
  {\bibfnamefont {G.}~\bibnamefont {Fischer}}, \ and\ \bibinfo {author}
  {\bibfnamefont {H.~v.}\ \bibnamefont {Löhneysen}},\ }\href {\doibase
  10.1038/srep42982} {\bibfield  {journal} {\bibinfo  {journal} {Scientific
  Reports}\ }\textbf {\bibinfo {volume} {7}},\ \bibinfo {pages} {42982}
  (\bibinfo {year} {2017})}\BibitemShut {NoStop}%
\bibitem [{\citenamefont {Luccas}\ \emph {et~al.}(2019)\citenamefont {Luccas},
  \citenamefont {Sánchez-Santolino}, \citenamefont {Correa-Orellana},
  \citenamefont {Mompean}, \citenamefont {García-Hernández},\ and\
  \citenamefont {Suderow}}]{luccas_magnetic_2019}%
  \BibitemOpen
  \bibfield  {author} {\bibinfo {author} {\bibfnamefont {R.~F.}\ \bibnamefont
  {Luccas}}, \bibinfo {author} {\bibfnamefont {G.}~\bibnamefont
  {Sánchez-Santolino}}, \bibinfo {author} {\bibfnamefont {A.}~\bibnamefont
  {Correa-Orellana}}, \bibinfo {author} {\bibfnamefont {F.~J.}\ \bibnamefont
  {Mompean}}, \bibinfo {author} {\bibfnamefont {M.}~\bibnamefont
  {García-Hernández}}, \ and\ \bibinfo {author} {\bibfnamefont
  {H.}~\bibnamefont {Suderow}},\ }\href {\doibase 10.1016/j.jmmm.2019.165451}
  {\bibfield  {journal} {\bibinfo  {journal} {Journal of Magnetism and Magnetic
  Materials}\ }\textbf {\bibinfo {volume} {489}},\ \bibinfo {pages} {165451}
  (\bibinfo {year} {2019})}\BibitemShut {NoStop}%
\bibitem [{\citenamefont {Gottschilch}\ \emph {et~al.}(2012)\citenamefont
  {Gottschilch}, \citenamefont {Gourdon}, \citenamefont {Persson},
  \citenamefont {Cruz}, \citenamefont {Petricek},\ and\ \citenamefont
  {Brueckel}}]{gottschilch_study_2012}%
  \BibitemOpen
  \bibfield  {author} {\bibinfo {author} {\bibfnamefont {M.}~\bibnamefont
  {Gottschilch}}, \bibinfo {author} {\bibfnamefont {O.}~\bibnamefont
  {Gourdon}}, \bibinfo {author} {\bibfnamefont {J.}~\bibnamefont {Persson}},
  \bibinfo {author} {\bibfnamefont {C.~d.~l.}\ \bibnamefont {Cruz}}, \bibinfo
  {author} {\bibfnamefont {V.}~\bibnamefont {Petricek}}, \ and\ \bibinfo
  {author} {\bibfnamefont {T.}~\bibnamefont {Brueckel}},\ }\href {\doibase
  10.1039/C2JM00154C} {\bibfield  {journal} {\bibinfo  {journal} {Journal of
  Materials Chemistry}\ }\textbf {\bibinfo {volume} {22}},\ \bibinfo {pages}
  {15275} (\bibinfo {year} {2012})},\ \bibinfo {note} {publisher: The Royal
  Society of Chemistry}\BibitemShut {NoStop}%
\bibitem [{\citenamefont {Brown}\ and\ \citenamefont
  {Forsyth}(1995)}]{brown_antiferromagnetism_1995}%
  \BibitemOpen
  \bibfield  {author} {\bibinfo {author} {\bibfnamefont {P.~J.}\ \bibnamefont
  {Brown}}\ and\ \bibinfo {author} {\bibfnamefont {J.~B.}\ \bibnamefont
  {Forsyth}},\ }\href {\doibase 10.1088/0953-8984/7/39/004} {\bibfield
  {journal} {\bibinfo  {journal} {Journal of Physics: Condensed Matter}\
  }\textbf {\bibinfo {volume} {7}},\ \bibinfo {pages} {7619} (\bibinfo {year}
  {1995})}\BibitemShut {NoStop}%
\bibitem [{\citenamefont {Lander}\ \emph {et~al.}(1967)\citenamefont {Lander},
  \citenamefont {Brown},\ and\ \citenamefont {Forsyth}}]{Lander_1967}%
  \BibitemOpen
  \bibfield  {author} {\bibinfo {author} {\bibfnamefont {G.~H.}\ \bibnamefont
  {Lander}}, \bibinfo {author} {\bibfnamefont {P.~J.}\ \bibnamefont {Brown}}, \
  and\ \bibinfo {author} {\bibfnamefont {J.~B.}\ \bibnamefont {Forsyth}},\
  }\href {\doibase 10.1088/0370-1328/91/2/310} {\bibfield  {journal} {\bibinfo
  {journal} {Proceedings of the Physical Society}\ }\textbf {\bibinfo {volume}
  {91}},\ \bibinfo {pages} {332} (\bibinfo {year} {1967})}\BibitemShut
  {NoStop}%
\bibitem [{\citenamefont {Menshikov}\ \emph {et~al.}(1990)\citenamefont
  {Menshikov}, \citenamefont {Vokhmyanin},\ and\ \citenamefont
  {Dorofeev}}]{Menshikov_1990}%
  \BibitemOpen
  \bibfield  {author} {\bibinfo {author} {\bibfnamefont {A.~Z.}\ \bibnamefont
  {Menshikov}}, \bibinfo {author} {\bibfnamefont {A.~P.}\ \bibnamefont
  {Vokhmyanin}}, \ and\ \bibinfo {author} {\bibfnamefont {Y.~A.}\ \bibnamefont
  {Dorofeev}},\ }\href {\doibase https://doi.org/10.1002/pssb.2221580132}
  {\bibfield  {journal} {\bibinfo  {journal} {physica status solidi (b)}\
  }\textbf {\bibinfo {volume} {158}},\ \bibinfo {pages} {319} (\bibinfo {year}
  {1990})}\BibitemShut {NoStop}%
\bibitem [{\citenamefont {Brown}\ \emph {et~al.}(1992)\citenamefont {Brown},
  \citenamefont {Forsyth}, \citenamefont {Nunez},\ and\ \citenamefont
  {Tasset}}]{brown_low-temperature_1992}%
  \BibitemOpen
  \bibfield  {author} {\bibinfo {author} {\bibfnamefont {P.~J.}\ \bibnamefont
  {Brown}}, \bibinfo {author} {\bibfnamefont {J.~B.}\ \bibnamefont {Forsyth}},
  \bibinfo {author} {\bibfnamefont {V.}~\bibnamefont {Nunez}}, \ and\ \bibinfo
  {author} {\bibfnamefont {F.}~\bibnamefont {Tasset}},\ }\href {\doibase
  10.1088/0953-8984/4/49/029} {\bibfield  {journal} {\bibinfo  {journal}
  {Journal of Physics: Condensed Matter}\ }\textbf {\bibinfo {volume} {4}},\
  \bibinfo {pages} {10025} (\bibinfo {year} {1992})}\BibitemShut {NoStop}%
\bibitem [{\citenamefont {dos Santos}\ \emph {et~al.}(2021)\citenamefont {dos
  Santos}, \citenamefont {Biniskos}, \citenamefont {Raymond}, \citenamefont
  {Schmalzl}, \citenamefont {dos Santos~Dias}, \citenamefont {Steffens},
  \citenamefont {Persson}, \citenamefont {Bl\"ugel}, \citenamefont {Lounis},\
  and\ \citenamefont {Br\"uckel}}]{dos_Santos_2021}%
  \BibitemOpen
  \bibfield  {author} {\bibinfo {author} {\bibfnamefont {F.~J.}\ \bibnamefont
  {dos Santos}}, \bibinfo {author} {\bibfnamefont {N.}~\bibnamefont
  {Biniskos}}, \bibinfo {author} {\bibfnamefont {S.}~\bibnamefont {Raymond}},
  \bibinfo {author} {\bibfnamefont {K.}~\bibnamefont {Schmalzl}}, \bibinfo
  {author} {\bibfnamefont {M.}~\bibnamefont {dos Santos~Dias}}, \bibinfo
  {author} {\bibfnamefont {P.}~\bibnamefont {Steffens}}, \bibinfo {author}
  {\bibfnamefont {J.}~\bibnamefont {Persson}}, \bibinfo {author} {\bibfnamefont
  {S.}~\bibnamefont {Bl\"ugel}}, \bibinfo {author} {\bibfnamefont
  {S.}~\bibnamefont {Lounis}}, \ and\ \bibinfo {author} {\bibfnamefont
  {T.}~\bibnamefont {Br\"uckel}},\ }\href {\doibase
  10.1103/PhysRevB.103.024407} {\bibfield  {journal} {\bibinfo  {journal}
  {Phys. Rev. B}\ }\textbf {\bibinfo {volume} {103}},\ \bibinfo {pages}
  {024407} (\bibinfo {year} {2021})}\BibitemShut {NoStop}%
\bibitem [{com({\natexlab{a}})}]{comment2}%
  \BibitemOpen
  \href@noop {} {} ({\natexlab{a}}),\ \bibinfo {note} {in the magnetic
  structure proposed by P.J. Brown et al., J. Phys.: Condens. Matter
  \textbf{4}, 10025 (1992) a discrepancy is noticed between the parameters of
  the magnetic moments given in the text and in the corresponding Table with
  the ones shown in the Figure of the same reference. Therefore, in our current
  work in Figure 1(b) we use the values of the reference given in the text and
  Table which we assume to be the correct ones.}\BibitemShut {Stop}%
\bibitem [{\citenamefont {Silva}\ \emph {et~al.}(2002)\citenamefont {Silva},
  \citenamefont {Brown},\ and\ \citenamefont {Forsyth}}]{silva_magnetic_2002}%
  \BibitemOpen
  \bibfield  {author} {\bibinfo {author} {\bibfnamefont {M.~R.}\ \bibnamefont
  {Silva}}, \bibinfo {author} {\bibfnamefont {P.~J.}\ \bibnamefont {Brown}}, \
  and\ \bibinfo {author} {\bibfnamefont {J.~B.}\ \bibnamefont {Forsyth}},\
  }\href {\doibase 10.1088/0953-8984/14/37/307} {\bibfield  {journal} {\bibinfo
   {journal} {Journal of Physics: Condensed Matter}\ }\textbf {\bibinfo
  {volume} {14}},\ \bibinfo {pages} {8707} (\bibinfo {year} {2002})},\ \bibinfo
  {note} {publisher: IOP Publishing}\BibitemShut {NoStop}%
\bibitem [{com({\natexlab{b}})}]{comment}%
  \BibitemOpen
  \href@noop {} {} ({\natexlab{b}}),\ \bibinfo {note} {a comment regarding the
  symmetry for the structure proposed by P.J. Brown et al., J. Phys.: Condens.
  Matter \textbf{4}, 10025 (1992) can be found in: MAGNDATA: A Collection of
  magnetic structures with portable cif-type files
  (\url{http://webbdcrista1.ehu.es/magndata/}) for
  $\textnormal{Mn}_5\textnormal{Si}_3$ \text{\#}1.307. We quote: ``The
  reference states that the structure is monoclinic, but in fact it has much
  lower symmetry (triclinic), with no symmetry except for the lattice and the
  antitranslation corresponding to the propagation vector. The spin
  correlations assumed in the model are not dictated by any possible
  symmetry.''}\BibitemShut {NoStop}%
\bibitem [{\citenamefont {Al-Kanani}\ and\ \citenamefont
  {Booth}(1995)}]{Alkanani_1995}%
  \BibitemOpen
  \bibfield  {author} {\bibinfo {author} {\bibfnamefont {H.}~\bibnamefont
  {Al-Kanani}}\ and\ \bibinfo {author} {\bibfnamefont {J.}~\bibnamefont
  {Booth}},\ }\href {\doibase https://doi.org/10.1016/0304-8853(94)01157-5}
  {\bibfield  {journal} {\bibinfo  {journal} {Journal of Magnetism and Magnetic
  Materials}\ }\textbf {\bibinfo {volume} {140-144}},\ \bibinfo {pages} {1539 }
  (\bibinfo {year} {1995})},\ \bibinfo {note} {international Conference on
  Magnetism}\BibitemShut {NoStop}%
\bibitem [{\citenamefont {Schmalzl}\ \emph {et~al.}(2016)\citenamefont
  {Schmalzl}, \citenamefont {Schmidt}, \citenamefont {Raymond}, \citenamefont
  {Feilbach}, \citenamefont {Mounier}, \citenamefont {Vettard},\ and\
  \citenamefont {Brückel}}]{schmalzl_upgrade_2016}%
  \BibitemOpen
  \bibfield  {author} {\bibinfo {author} {\bibfnamefont {K.}~\bibnamefont
  {Schmalzl}}, \bibinfo {author} {\bibfnamefont {W.}~\bibnamefont {Schmidt}},
  \bibinfo {author} {\bibfnamefont {S.}~\bibnamefont {Raymond}}, \bibinfo
  {author} {\bibfnamefont {H.}~\bibnamefont {Feilbach}}, \bibinfo {author}
  {\bibfnamefont {C.}~\bibnamefont {Mounier}}, \bibinfo {author} {\bibfnamefont
  {B.}~\bibnamefont {Vettard}}, \ and\ \bibinfo {author} {\bibfnamefont
  {T.}~\bibnamefont {Brückel}},\ }\href {\doibase 10.1016/j.nima.2016.02.067}
  {\bibfield  {journal} {\bibinfo  {journal} {Nuclear Instruments and Methods
  in Physics Research Section A: Accelerators, Spectrometers, Detectors and
  Associated Equipment}\ }\textbf {\bibinfo {volume} {819}},\ \bibinfo {pages}
  {89} (\bibinfo {year} {2016})}\BibitemShut {NoStop}%
\bibitem [{dat(2016)}]{data_IN12a}%
  \BibitemOpen
  \href {https://doi.ill.fr/10.5291/ILL-DATA.CRG-2331} {\bibfield  {journal}
  {\bibinfo  {journal} {https://doi.ill.fr/10.5291/ILL-DATA.CRG-2331}\ }
  (\bibinfo {year} {2016})}\BibitemShut {NoStop}%
\bibitem [{dat(2020)}]{data_IN12b}%
  \BibitemOpen
  \href {https://doi.ill.fr/10.5291/ILL-DATA.INTER-504} {\bibfield  {journal}
  {\bibinfo  {journal} {https://doi.ill.fr/10.5291/ILL-DATA.INTER-504}\ }
  (\bibinfo {year} {2020})}\BibitemShut {NoStop}%
\bibitem [{dat(2021)}]{data_IN22}%
  \BibitemOpen
  \href {https://doi.ill.fr/10.5291/ILL-DATA.CRG-2380} {\bibfield  {journal}
  {\bibinfo  {journal} {https://doi.ill.fr/10.5291/ILL-DATA.CRG-2380}\ }
  (\bibinfo {year} {2021})}\BibitemShut {NoStop}%
\bibitem [{\citenamefont {Papanikolaou}\ \emph {et~al.}(2002)\citenamefont
  {Papanikolaou}, \citenamefont {Zeller},\ and\ \citenamefont
  {Dederichs}}]{papanikolaou_conceptual_2002}%
  \BibitemOpen
  \bibfield  {author} {\bibinfo {author} {\bibfnamefont {N.}~\bibnamefont
  {Papanikolaou}}, \bibinfo {author} {\bibfnamefont {R.}~\bibnamefont
  {Zeller}}, \ and\ \bibinfo {author} {\bibfnamefont {P.~H.}\ \bibnamefont
  {Dederichs}},\ }\href {\doibase 10.1088/0953-8984/14/11/304} {\bibfield
  {journal} {\bibinfo  {journal} {Journal of Physics: Condensed Matter}\
  }\textbf {\bibinfo {volume} {14}},\ \bibinfo {pages} {2799} (\bibinfo {year}
  {2002})},\ \bibinfo {note} {publisher: IOP Publishing}\BibitemShut {NoStop}%
\bibitem [{\citenamefont {Vosko}\ \emph {et~al.}(1980)\citenamefont {Vosko},
  \citenamefont {Wilk},\ and\ \citenamefont {Nusair}}]{vosko_accurate_1980}%
  \BibitemOpen
  \bibfield  {author} {\bibinfo {author} {\bibfnamefont {S.~H.}\ \bibnamefont
  {Vosko}}, \bibinfo {author} {\bibfnamefont {L.}~\bibnamefont {Wilk}}, \ and\
  \bibinfo {author} {\bibfnamefont {M.}~\bibnamefont {Nusair}},\ }\href
  {\doibase 10.1139/p80-159} {\bibfield  {journal} {\bibinfo  {journal}
  {Canadian Journal of Physics}\ }\textbf {\bibinfo {volume} {58}},\ \bibinfo
  {pages} {1200} (\bibinfo {year} {1980})}\BibitemShut {NoStop}%
\bibitem [{\citenamefont {Liechtenstein}\ \emph {et~al.}(1987)\citenamefont
  {Liechtenstein}, \citenamefont {Katsnelson}, \citenamefont {Antropov},\ and\
  \citenamefont {Gubanov}}]{liechtenstein_local_1987}%
  \BibitemOpen
  \bibfield  {author} {\bibinfo {author} {\bibfnamefont {A.~I.}\ \bibnamefont
  {Liechtenstein}}, \bibinfo {author} {\bibfnamefont {M.~I.}\ \bibnamefont
  {Katsnelson}}, \bibinfo {author} {\bibfnamefont {V.~P.}\ \bibnamefont
  {Antropov}}, \ and\ \bibinfo {author} {\bibfnamefont {V.~A.}\ \bibnamefont
  {Gubanov}},\ }\href {\doibase 10.1016/0304-8853(87)90721-9} {\bibfield
  {journal} {\bibinfo  {journal} {Journal of Magnetism and Magnetic Materials}\
  }\textbf {\bibinfo {volume} {67}},\ \bibinfo {pages} {65} (\bibinfo {year}
  {1987})}\BibitemShut {NoStop}%
\bibitem [{\citenamefont {Ebert}\ and\ \citenamefont
  {Mankovsky}(2009)}]{ebert_anisotropic_2009}%
  \BibitemOpen
  \bibfield  {author} {\bibinfo {author} {\bibfnamefont {H.}~\bibnamefont
  {Ebert}}\ and\ \bibinfo {author} {\bibfnamefont {S.}~\bibnamefont
  {Mankovsky}},\ }\href {\doibase 10.1103/PhysRevB.79.045209} {\bibfield
  {journal} {\bibinfo  {journal} {Physical Review B}\ }\textbf {\bibinfo
  {volume} {79}},\ \bibinfo {pages} {045209} (\bibinfo {year}
  {2009})}\BibitemShut {NoStop}%
\bibitem [{\citenamefont {M\"uller}\ \emph {et~al.}(2019)\citenamefont
  {M\"uller}, \citenamefont {Hoffmann}, \citenamefont {Di{\ss}elkamp},
  \citenamefont {Schürhoff}, \citenamefont {Mavros}, \citenamefont
  {Sallermann}, \citenamefont {Kiselev}, \citenamefont {J\'onsson},\ and\
  \citenamefont {Bl\"ugel}}]{muller_spirit:_2019}%
  \BibitemOpen
  \bibfield  {author} {\bibinfo {author} {\bibfnamefont {G.~P.}\ \bibnamefont
  {M\"uller}}, \bibinfo {author} {\bibfnamefont {M.}~\bibnamefont {Hoffmann}},
  \bibinfo {author} {\bibfnamefont {C.}~\bibnamefont {Di{\ss}elkamp}}, \bibinfo
  {author} {\bibfnamefont {D.}~\bibnamefont {Schürhoff}}, \bibinfo {author}
  {\bibfnamefont {S.}~\bibnamefont {Mavros}}, \bibinfo {author} {\bibfnamefont
  {M.}~\bibnamefont {Sallermann}}, \bibinfo {author} {\bibfnamefont {N.~S.}\
  \bibnamefont {Kiselev}}, \bibinfo {author} {\bibfnamefont {H.}~\bibnamefont
  {J\'onsson}}, \ and\ \bibinfo {author} {\bibfnamefont {S.}~\bibnamefont
  {Bl\"ugel}},\ }\href {\doibase 10.1103/PhysRevB.99.224414} {\bibfield
  {journal} {\bibinfo  {journal} {Physical Review B}\ }\textbf {\bibinfo
  {volume} {99}},\ \bibinfo {pages} {224414} (\bibinfo {year}
  {2019})}\BibitemShut {NoStop}%
\bibitem [{\citenamefont {dos Santos}\ \emph {et~al.}(2018)\citenamefont {dos
  Santos}, \citenamefont {dos Santos~Dias}, \citenamefont {Guimarães},
  \citenamefont {Bouaziz},\ and\ \citenamefont
  {Lounis}}]{dos_santos_spin-resolved_2018}%
  \BibitemOpen
  \bibfield  {author} {\bibinfo {author} {\bibfnamefont {F.~J.}\ \bibnamefont
  {dos Santos}}, \bibinfo {author} {\bibfnamefont {M.}~\bibnamefont {dos
  Santos~Dias}}, \bibinfo {author} {\bibfnamefont {F.~S.~M.}\ \bibnamefont
  {Guimarães}}, \bibinfo {author} {\bibfnamefont {J.}~\bibnamefont {Bouaziz}},
  \ and\ \bibinfo {author} {\bibfnamefont {S.}~\bibnamefont {Lounis}},\ }\href
  {\doibase 10.1103/PhysRevB.97.024431} {\bibfield  {journal} {\bibinfo
  {journal} {Physical Review B}\ }\textbf {\bibinfo {volume} {97}},\ \bibinfo
  {pages} {024431} (\bibinfo {year} {2018})}\BibitemShut {NoStop}%
\bibitem [{\citenamefont {dos Santos}\ \emph
  {et~al.}(2020{\natexlab{a}})\citenamefont {dos Santos}, \citenamefont {dos
  Santos~Dias},\ and\ \citenamefont {Lounis}}]{dos_santos_nonreciprocity_2020}%
  \BibitemOpen
  \bibfield  {author} {\bibinfo {author} {\bibfnamefont {F.~J.}\ \bibnamefont
  {dos Santos}}, \bibinfo {author} {\bibfnamefont {M.}~\bibnamefont {dos
  Santos~Dias}}, \ and\ \bibinfo {author} {\bibfnamefont {S.}~\bibnamefont
  {Lounis}},\ }\href {\doibase 10.1103/PhysRevB.102.104401} {\bibfield
  {journal} {\bibinfo  {journal} {Physical Review B}\ }\textbf {\bibinfo
  {volume} {102}},\ \bibinfo {pages} {104401} (\bibinfo {year}
  {2020}{\natexlab{a}})},\ \bibinfo {note} {publisher: American Physical
  Society}\BibitemShut {NoStop}%
\bibitem [{\citenamefont {dos Santos}\ \emph
  {et~al.}(2020{\natexlab{b}})\citenamefont {dos Santos}, \citenamefont {dos
  Santos~Dias},\ and\ \citenamefont {Lounis}}]{dos_santos_modeling_2020}%
  \BibitemOpen
  \bibfield  {author} {\bibinfo {author} {\bibfnamefont {F.~J.}\ \bibnamefont
  {dos Santos}}, \bibinfo {author} {\bibfnamefont {M.}~\bibnamefont {dos
  Santos~Dias}}, \ and\ \bibinfo {author} {\bibfnamefont {S.}~\bibnamefont
  {Lounis}},\ }\href {\doibase 10.1103/PhysRevB.102.104436} {\bibfield
  {journal} {\bibinfo  {journal} {Physical Review B}\ }\textbf {\bibinfo
  {volume} {102}},\ \bibinfo {pages} {104436} (\bibinfo {year}
  {2020}{\natexlab{b}})},\ \bibinfo {note} {publisher: American Physical
  Society}\BibitemShut {NoStop}%
\bibitem [{\citenamefont {Ibuka}\ \emph {et~al.}(2017)\citenamefont {Ibuka},
  \citenamefont {Itoh}, \citenamefont {Yokoo},\ and\ \citenamefont
  {Endoh}}]{Ibuka}%
  \BibitemOpen
  \bibfield  {author} {\bibinfo {author} {\bibfnamefont {S.}~\bibnamefont
  {Ibuka}}, \bibinfo {author} {\bibfnamefont {S.}~\bibnamefont {Itoh}},
  \bibinfo {author} {\bibfnamefont {T.}~\bibnamefont {Yokoo}}, \ and\ \bibinfo
  {author} {\bibfnamefont {Y.}~\bibnamefont {Endoh}},\ }\href {\doibase
  10.1103/PhysRevB.95.224406} {\bibfield  {journal} {\bibinfo  {journal} {Phys.
  Rev. B}\ }\textbf {\bibinfo {volume} {95}},\ \bibinfo {pages} {224406}
  (\bibinfo {year} {2017})}\BibitemShut {NoStop}%
\bibitem [{\citenamefont {Diallo}\ \emph {et~al.}(2009)\citenamefont {Diallo},
  \citenamefont {Antropov}, \citenamefont {Perring}, \citenamefont {Broholm},
  \citenamefont {Pulikkotil}, \citenamefont {Ni}, \citenamefont {Bud'ko},
  \citenamefont {Canfield}, \citenamefont {Kreyssig}, \citenamefont {Goldman},\
  and\ \citenamefont {McQueeney}}]{Diallo}%
  \BibitemOpen
  \bibfield  {author} {\bibinfo {author} {\bibfnamefont {S.~O.}\ \bibnamefont
  {Diallo}}, \bibinfo {author} {\bibfnamefont {V.~P.}\ \bibnamefont
  {Antropov}}, \bibinfo {author} {\bibfnamefont {T.~G.}\ \bibnamefont
  {Perring}}, \bibinfo {author} {\bibfnamefont {C.}~\bibnamefont {Broholm}},
  \bibinfo {author} {\bibfnamefont {J.~J.}\ \bibnamefont {Pulikkotil}},
  \bibinfo {author} {\bibfnamefont {N.}~\bibnamefont {Ni}}, \bibinfo {author}
  {\bibfnamefont {S.~L.}\ \bibnamefont {Bud'ko}}, \bibinfo {author}
  {\bibfnamefont {P.~C.}\ \bibnamefont {Canfield}}, \bibinfo {author}
  {\bibfnamefont {A.}~\bibnamefont {Kreyssig}}, \bibinfo {author}
  {\bibfnamefont {A.~I.}\ \bibnamefont {Goldman}}, \ and\ \bibinfo {author}
  {\bibfnamefont {R.~J.}\ \bibnamefont {McQueeney}},\ }\href {\doibase
  10.1103/PhysRevLett.102.187206} {\bibfield  {journal} {\bibinfo  {journal}
  {Phys. Rev. Lett.}\ }\textbf {\bibinfo {volume} {102}},\ \bibinfo {pages}
  {187206} (\bibinfo {year} {2009})}\BibitemShut {NoStop}%
\bibitem [{\citenamefont {Harriger}\ \emph {et~al.}(2011)\citenamefont
  {Harriger}, \citenamefont {Luo}, \citenamefont {Liu}, \citenamefont {Frost},
  \citenamefont {Hu}, \citenamefont {Norman},\ and\ \citenamefont
  {Dai}}]{Harriger_2011}%
  \BibitemOpen
  \bibfield  {author} {\bibinfo {author} {\bibfnamefont {L.~W.}\ \bibnamefont
  {Harriger}}, \bibinfo {author} {\bibfnamefont {H.~Q.}\ \bibnamefont {Luo}},
  \bibinfo {author} {\bibfnamefont {M.~S.}\ \bibnamefont {Liu}}, \bibinfo
  {author} {\bibfnamefont {C.}~\bibnamefont {Frost}}, \bibinfo {author}
  {\bibfnamefont {J.~P.}\ \bibnamefont {Hu}}, \bibinfo {author} {\bibfnamefont
  {M.~R.}\ \bibnamefont {Norman}}, \ and\ \bibinfo {author} {\bibfnamefont
  {P.}~\bibnamefont {Dai}},\ }\href {\doibase 10.1103/PhysRevB.84.054544}
  {\bibfield  {journal} {\bibinfo  {journal} {Phys. Rev. B}\ }\textbf {\bibinfo
  {volume} {84}},\ \bibinfo {pages} {054544} (\bibinfo {year}
  {2011})}\BibitemShut {NoStop}%
\bibitem [{\citenamefont {Wang}\ \emph {et~al.}(2015)\citenamefont {Wang},
  \citenamefont {Valdivia}, \citenamefont {Yi}, \citenamefont {Chen},
  \citenamefont {Zhang}, \citenamefont {Ewings}, \citenamefont {Perring},
  \citenamefont {Zhao}, \citenamefont {Harriger}, \citenamefont {Lynn},
  \citenamefont {Bourret-Courchesne}, \citenamefont {Dai}, \citenamefont {Lee},
  \citenamefont {Yao},\ and\ \citenamefont {Birgeneau}}]{Wang_2015}%
  \BibitemOpen
  \bibfield  {author} {\bibinfo {author} {\bibfnamefont {M.}~\bibnamefont
  {Wang}}, \bibinfo {author} {\bibfnamefont {P.}~\bibnamefont {Valdivia}},
  \bibinfo {author} {\bibfnamefont {M.}~\bibnamefont {Yi}}, \bibinfo {author}
  {\bibfnamefont {J.~X.}\ \bibnamefont {Chen}}, \bibinfo {author}
  {\bibfnamefont {W.~L.}\ \bibnamefont {Zhang}}, \bibinfo {author}
  {\bibfnamefont {R.~A.}\ \bibnamefont {Ewings}}, \bibinfo {author}
  {\bibfnamefont {T.~G.}\ \bibnamefont {Perring}}, \bibinfo {author}
  {\bibfnamefont {Y.}~\bibnamefont {Zhao}}, \bibinfo {author} {\bibfnamefont
  {L.~W.}\ \bibnamefont {Harriger}}, \bibinfo {author} {\bibfnamefont {J.~W.}\
  \bibnamefont {Lynn}}, \bibinfo {author} {\bibfnamefont {E.}~\bibnamefont
  {Bourret-Courchesne}}, \bibinfo {author} {\bibfnamefont {P.}~\bibnamefont
  {Dai}}, \bibinfo {author} {\bibfnamefont {D.~H.}\ \bibnamefont {Lee}},
  \bibinfo {author} {\bibfnamefont {D.~X.}\ \bibnamefont {Yao}}, \ and\
  \bibinfo {author} {\bibfnamefont {R.~J.}\ \bibnamefont {Birgeneau}},\ }\href
  {\doibase 10.1103/PhysRevB.92.041109} {\bibfield  {journal} {\bibinfo
  {journal} {Phys. Rev. B}\ }\textbf {\bibinfo {volume} {92}},\ \bibinfo
  {pages} {041109} (\bibinfo {year} {2015})}\BibitemShut {NoStop}%
\bibitem [{\citenamefont {Jacobsen}\ \emph {et~al.}(2018)\citenamefont
  {Jacobsen}, \citenamefont {Gaw}, \citenamefont {Princep}, \citenamefont
  {Hamilton}, \citenamefont {T\'oth}, \citenamefont {Ewings}, \citenamefont
  {Enderle}, \citenamefont {Wheeler}, \citenamefont {Prabhakaran},\ and\
  \citenamefont {Boothroyd}}]{Jacobsen_2018}%
  \BibitemOpen
  \bibfield  {author} {\bibinfo {author} {\bibfnamefont {H.}~\bibnamefont
  {Jacobsen}}, \bibinfo {author} {\bibfnamefont {S.~M.}\ \bibnamefont {Gaw}},
  \bibinfo {author} {\bibfnamefont {A.~J.}\ \bibnamefont {Princep}}, \bibinfo
  {author} {\bibfnamefont {E.}~\bibnamefont {Hamilton}}, \bibinfo {author}
  {\bibfnamefont {S.}~\bibnamefont {T\'oth}}, \bibinfo {author} {\bibfnamefont
  {R.~A.}\ \bibnamefont {Ewings}}, \bibinfo {author} {\bibfnamefont
  {M.}~\bibnamefont {Enderle}}, \bibinfo {author} {\bibfnamefont {E.~M.~H.}\
  \bibnamefont {Wheeler}}, \bibinfo {author} {\bibfnamefont {D.}~\bibnamefont
  {Prabhakaran}}, \ and\ \bibinfo {author} {\bibfnamefont {A.~T.}\ \bibnamefont
  {Boothroyd}},\ }\href {\doibase 10.1103/PhysRevB.97.144401} {\bibfield
  {journal} {\bibinfo  {journal} {Phys. Rev. B}\ }\textbf {\bibinfo {volume}
  {97}},\ \bibinfo {pages} {144401} (\bibinfo {year} {2018})}\BibitemShut
  {NoStop}%
\bibitem [{\citenamefont {Wuttke}\ \emph {et~al.}(2019)\citenamefont {Wuttke},
  \citenamefont {Caglieris}, \citenamefont {Sykora}, \citenamefont
  {Scaravaggi}, \citenamefont {Wolter}, \citenamefont {Manna}, \citenamefont
  {S\"uss}, \citenamefont {Shekhar}, \citenamefont {Felser}, \citenamefont
  {B\"uchner},\ and\ \citenamefont {Hess}}]{Wuttke_2019}%
  \BibitemOpen
  \bibfield  {author} {\bibinfo {author} {\bibfnamefont {C.}~\bibnamefont
  {Wuttke}}, \bibinfo {author} {\bibfnamefont {F.}~\bibnamefont {Caglieris}},
  \bibinfo {author} {\bibfnamefont {S.}~\bibnamefont {Sykora}}, \bibinfo
  {author} {\bibfnamefont {F.}~\bibnamefont {Scaravaggi}}, \bibinfo {author}
  {\bibfnamefont {A.~U.~B.}\ \bibnamefont {Wolter}}, \bibinfo {author}
  {\bibfnamefont {K.}~\bibnamefont {Manna}}, \bibinfo {author} {\bibfnamefont
  {V.}~\bibnamefont {S\"uss}}, \bibinfo {author} {\bibfnamefont
  {C.}~\bibnamefont {Shekhar}}, \bibinfo {author} {\bibfnamefont
  {C.}~\bibnamefont {Felser}}, \bibinfo {author} {\bibfnamefont
  {B.}~\bibnamefont {B\"uchner}}, \ and\ \bibinfo {author} {\bibfnamefont
  {C.}~\bibnamefont {Hess}},\ }\href {\doibase 10.1103/PhysRevB.100.085111}
  {\bibfield  {journal} {\bibinfo  {journal} {Phys. Rev. B}\ }\textbf {\bibinfo
  {volume} {100}},\ \bibinfo {pages} {085111} (\bibinfo {year}
  {2019})}\BibitemShut {NoStop}%
\bibitem [{\citenamefont {Xiao}\ \emph {et~al.}(2006)\citenamefont {Xiao},
  \citenamefont {Yao}, \citenamefont {Fang},\ and\ \citenamefont
  {Niu}}]{Xiao_2006}%
  \BibitemOpen
  \bibfield  {author} {\bibinfo {author} {\bibfnamefont {D.}~\bibnamefont
  {Xiao}}, \bibinfo {author} {\bibfnamefont {Y.}~\bibnamefont {Yao}}, \bibinfo
  {author} {\bibfnamefont {Z.}~\bibnamefont {Fang}}, \ and\ \bibinfo {author}
  {\bibfnamefont {Q.}~\bibnamefont {Niu}},\ }\href {\doibase
  10.1103/PhysRevLett.97.026603} {\bibfield  {journal} {\bibinfo  {journal}
  {Phys. Rev. Lett.}\ }\textbf {\bibinfo {volume} {97}},\ \bibinfo {pages}
  {026603} (\bibinfo {year} {2006})}\BibitemShut {NoStop}%
\end{thebibliography}%


%

\end{document}